\documentclass[aps,pre,amsmath,amssymb,reprint,superscriptaddress]{revtex4-1}

\usepackage{graphicx}
\usepackage{subcaption}
\usepackage{sidecap}
\usepackage{hyperref}
\hypersetup{colorlinks,}

\newcommand{\Nm}{N_{\mathrm{m}}}

\newcommand{\rc}{r_{\mathrm{c}}}
\newcommand{\kTc}{k_{\mathrm{B}}T_{\mathrm{c}}}
\newcommand{\kBT}{k_{\mathrm{B}}T}

\newcommand{\rd}{r_{\mathrm{d}}}
\newcommand{\wwd}{w_{\mathrm{d}}}

\newcommand{\vc}[1]{\mathbf{ #1}}

\begin{document}
\title{Liquid phase parametrisation and solidification in many-body dissipative particle dynamics}

\author{Peter Vanya}
\email{peter.vanya@gmail.com}
\affiliation{Department of Materials Science \& Metallurgy, University of Cambridge, 27 Charles Babbage Road, Cambridge CB3 0FS, United Kingdom}

\author{Phillip Crout}
\affiliation{Department of Materials Science \& Metallurgy, University of Cambridge, 27 Charles Babbage Road, Cambridge CB3 0FS, United Kingdom}

\author{Jonathan Sharman}
\affiliation{Johnson Matthey Technology Centre, Blounts Court Road, Sonning Common, Reading RG4 9NH, United Kingdom}

\author{James A. Elliott}
\email{jae1001@cam.ac.uk}
\affiliation{Department of Materials Science \& Metallurgy, University of Cambridge, 27 Charles Babbage Road, Cambridge CB3 0FS, United Kingdom}

\date{\today}

\begin{abstract}
Many-body dissipative particle dynamics (MDPD) is a mesoscale method capable of reproducing liquid-vapour coexistence in a single simulation. Despite having been introduced more than a decade ago, this method remains broadly unexplored and, as a result, relatively unused for modelling of industrially important soft matter systems. In this work, we systematically investigate the structure and properties of an MDPD fluid. We show that, besides the liquid phase, the MDPD potential can also yield a gas phase and a thermodynamically stable solid phase with a bcc lattice, but lacking a proper stress-strain relation. For the liquid phase, we determine the dependence of density and surface tension on the interaction parameters, and devise a top-down parametrisation protocol for real liquids.
\end{abstract}

\maketitle

\section{Introduction}
In designing a new force field it is vital to understand its phase diagram before applying it to real systems. It is generally prohibitively expensive to derive the equation of state (EOS), from which all the experimental observables would follow, from a molecular dynamics force field, due to many parameters that can be varied. However, the EOS can be inferred for some mesoscale potentials, which possess few parameters. This is the case for dissipative particle dynamics (DPD), for which the EOS can be easily reverse-engineered.

The standard DPD method was introduced by Hoogerbrugge and Koelman~\cite{Hoogerbrugge_EPL_1992} and thoroughly explored by Groot and Warren~\cite{Groot_JCP_1997}, who derived the EOS and linked it to the Flory-Huggins theory for polymer mixtures. It was consequently shown that this method can reproduce diblock copolymer phases~\cite{Groot_JCP_1998}, vesicle formation~\cite{Yamamoto_JCP_2002}, or the morphology of ionomer membranes~\citep{Yamamoto_PolymJ_2003,Wu_EES_2008}, among many other soft matter systems. Over the past 20 years, DPD has become an important tool in gaining insight into soft matter structures on the scale of 1-100 nm~\cite{Espanol_JCP_2017}.

However, the standard DPD method has an important drawback in that its purely repulsive force field:
\begin{equation}
F(r) = \begin{cases}
A(1-r),& r<1,\\
0,& r\geq 1\\
\end{cases}
\end{equation}
with $r$ being distance between two particles and parameter $A>0$, cannot support liquid-vapour coexistence. In order to overcome this deficiency and retain the simplicity and other advantages of the extremely soft potential, there have been several attempts to generalise DPD and increase its scope of applicability. A simple extension called many-body DPD (MDPD) adds a density-dependent repulsive term\cite{Pagonabarraga_JCP_2001,Trofimov_JCP_2002,Warren_PRE_2003}
\begin{equation}
F_{\rm{rep}}(r)=\begin{cases}
B(\bar\rho_i+\bar\rho_j)(1-r/\rd),& r<\rd,\\
0,& r\geq \rd,\\
\end{cases}
\end{equation}
where $B>0$ is the repulsion parameter, $\rd<1$ a new, many-body lengthscale, and $\bar\rho_i$ a local density for $i$th particle (defined below in eq.~\eqref{eq:rho_loc}). For some specific set of parameters, this force field can simulate a water slab with correct surface tension~\cite{Ghoufi_PRE_2011}. Since its introduction, MDPD has been linked to Flory-Huggins theory~\cite{Jamali_JCP_2015, Ghoufi_JCTC_2012} and tested on several simplified models of pure liquids~\cite{Ghoufi_JCTC_2012, Ghoufi_EPJE_2013, Atashafrooz_JCED_2016} or polymers~\cite{Yong_Polymers_2016}. However, the scope of its applications is still limited, when compared with standard DPD, and so far this method has not been applied to more complex systems.

The first restriction on the applicability of MDPD is the lack of a systematic protocol for generating the interaction parameters that would reproduce the properties of real liquids. For example, Ghoufi \emph{et al.}~\cite{Ghoufi_PRE_2011} simulated pure water at a coarse-graining (CG) degree of three molecules per bead, and showed that their set of parameters leads to the correct density and surface tension. However, the authors did not explain how they generated these parameters, or how these should be modified if one wanted to simulate water at a different CG degree.

Secondly, while the behaviour of standard DPD, which is controlled by only one interaction parameter, $A$, is relatively well understood, MDPD has three: $A,B$ and $\rd$. The two additional parameters significantly increase the complexity of the phase diagram and the risk of unexpected and unphysical behaviour if not chosen well.

The aim of this paper is to resolve these two problems. To this end, we explore a large portion of the phase diagram of an MDPD fluid by systematically varying the values of repulsion $B$, attraction $A$ and many-body cutoff $\rd$. By measuring the density and the self-diffusion coefficient, we reveal the region of the liquid-vapour coexistence as well as the gas phase, where the particles homogeneously fill the whole simulation cell, and a solid phase with a well-defined lattice and negligible particle diffusion, but lacking a proper stress-strain relation. Having determined the phase boundaries, we then discuss how these findings can be applied to define a top-down parametrisation protocol. Finally, we demonstrate how this protocol can yield the interaction parameters for water at varying CG degrees.

We note that there is an extension generalising both DPD and MDPD called smoothed DPD (SDPD). This method corrects for the problems with transport and an inability to simulate non-isothermal settings based on discretising Navier-Stokes equations~\cite{Espanol_PRE_2003, Litvinov_PRE_2008}. However, the simplicity and versatility of MDPD makes the effort of parametrising it a worthwhile pursuit before considering a more general SDPD.

Section~\ref{sec:method} reviews the MDPD force field. In Section~\ref{sec:sim} we present tools used for determining the phase behaviour, namely the density profile, self-diffusivity, surface tension, and coordination number, and determine the lattice of the solid phase. In Section~\ref{sec:real}, we present the top-down parametrisation protocol for the liquid phase and derive the interaction parameters for a few solvents.

\section{The method}
\label{sec:method}
Adopting a set of reduced units such that particle size $\rc=1$, mass $m=1$ and temperature $\kBT=1$ in the spirit of the original DPD paper~\cite{Groot_JCP_1997}, the full form of the MDPD force field is:
\begin{equation}
\vc F_{ij}(\vc r) = A w(r) \vc{\hat r} + B(\bar\rho_i + \bar\rho_j) \wwd (r) \vc{\hat r},
\label{eq:ff}
\end{equation}
where $A$ and $B$ are interaction parameters, $r=|\vc r|$, the weight functions are:
\begin{equation}
w(r)=\begin{cases}
1-r,& r<1,\\
0,& r \geq 1,
\end{cases}
\end{equation}
\begin{equation}
\wwd(r)=\begin{cases}
1-r/\rd,& r<\rd,\\
0,& r \geq \rd,
\end{cases}
\end{equation}
and the local density $\bar\rho_i$ around particle $i$ is defined as:
\begin{equation}
\bar \rho_i = \sum_{j\neq i} \frac {15} {2\pi \rd^3} \wwd(r_{ij})^2,
\label{eq:rho_loc}
\end{equation}

Warren showed that for $A<0$ and $B>0$ this force field leads to the liquid-vapour coexistence, and derived the EOS~\cite{Warren_PRE_2003}:
\begin{equation}
p = \rho\kBT + \alpha A \rho^2 + 2\alpha B \rd^4 (\rho^3 - c\rho^2 + d),
\label{eq:warren}
\end{equation}
with fitting constants $\alpha=0.1, c=4.16$, and $d=18$. This EOS was revisited by Jamali~\cite{Jamali_JCP_2015}, who came with a slightly different and arguably more precise form:
\begin{equation}
p = \rho\kBT + \alpha A\rho^2 + 2\alpha B \rd^4 (\rho^3 - c' \rho^2 + d' \rho) 
-\frac{\alpha B\rd^4}{|A|^{1/2}}\rho^2,
\label{eq:jamali}
\end{equation}
where $c'=4.69$ and $d'=7.55$. In practice, the difference between these two EOS's is small for typical liquid densities, \emph{e.g.} at $A=-40,B=25,\rho=6$ it is about 5\%.

In the simulation, the system is thermostatted by the DPD thermostat introduced by Espa\~{n}ol and Warren~\cite{Espanol_EPL_1995} via the dissipative and random force:
\begin{align}
F^{\rm D}(\vc r) &= -\gamma\, w(r) (\vc v \cdot \vc{\hat r}) \vc{\hat r},\\
F^{\rm R}(\vc r) &= \sqrt{2\gamma\kBT} w^2(r) \frac{\theta}{\sqrt{\Delta t}} \vc{\hat r},
\end{align}
where $\gamma$ is the friction parameter, $\theta$ is a Gaussian random number with zero mean and unit variance, and $\Delta t$ the simulation timestep.

In the standard DPD, the simulation density is decided \emph{a priori}, and most often is equal to 3, which is the lowest possible number at which the EOS is still quadratic. This value then remains fixed throughout the simulation by the constraint of constant volume. However, the density in an MDPD liquid can arise naturally by choosing the right interaction parameters $A$, $B$ and $\rd$ at which the liquid forms a droplet with a surface. In this sense, it resembles a classical molecular dynamics force field.

In varying $A,B,\rd$ there are several obvious constraints. Firstly, we choose $0<\rd<1$, $A<0$, $B>0$ to make the interaction attractive near $r=1$ and repulsive at the core near $r=0$. In fact, to ensure that $F(0)>0$, it follows from eq.~\eqref{eq:ff} that $B > -A 2\pi\rd^3 / 15$. Even values close to this boundary might lead to poor temperature conservation. Henceforth we will call this a \emph{no-go} region.

\subsection{Simulation details}
\label{sec:sim}
Following Ghoufi \emph{et al.}~\cite{Ghoufi_PRE_2011}, we set a simulation cell size of $22\times 5\times 5$ (in reduced units), with one dimension significantly larger than the others. This asymmetry forces the liquid to form a rectangular slab instead of a spherical droplet, which facilitates calculation of the surface tension. The simulation step $\Delta t$ is set to 0.01, which is significantly lower than the one used in standard DPD simulations (0.05). The MDPD force field is not strictly linear and so one should expect the need to lower the simulation step in order to keep the temperature within manageable limits. On several occasions, especially at lower values of $\rd$, the temperature in our simulations diverged by more than 10\%, which is considered undesirable~\cite{Warren_PRE_2003}. While this problem can be generally ameliorated by further lowering the timestep, this creates a penalty in the form of decreased simulation speed and undermines the main advantage of DPD and MDPD as a mesoscale method. For this reason, we did not use timesteps below 0.01 and did not explore many-body cutoffs below $\rd=0.55$. 

In each simulation we used 1000 particles, equilibrated for 500k steps and measured during the following 5k steps, a long enough interval to capture mass transport since a bead with average speed would be displaced by 50 length units. The dissipation parameter $\gamma$ was set to 4.5, a value commonly used in the literature. We note that varying $\gamma$ would change the diffusive behaviour, but it would not influence the position of phase boundaries or equilibrium behaviour in general. To perform the simulations we used the DL\_MESO software package version 2.6~\cite{DLMS}.

We have explored a wide range of values $A$ and $B$. We also note that $A$ should always be negative in order to create van der Waals loop~\cite{Warren_PRE_2003} and the liquid-vapour interface, whereas values of $B$ should always be positive to keep the core of the force field repulsive. We chose the range $[-100, 0]$ for $A$ and $[0, 100]$ for $B$ and henceforth refer to them as attraction and repulsion, respectively. In Section~\ref{sec:real}, we will show that a real liquid can fall into this range for a wide number of CG degrees.

\section{Measurement of properties}
\subsection{Density}
\begin{figure*}
\centering
\begin{subfigure}[b]{0.31\textwidth}
\includegraphics[width=1\textwidth]{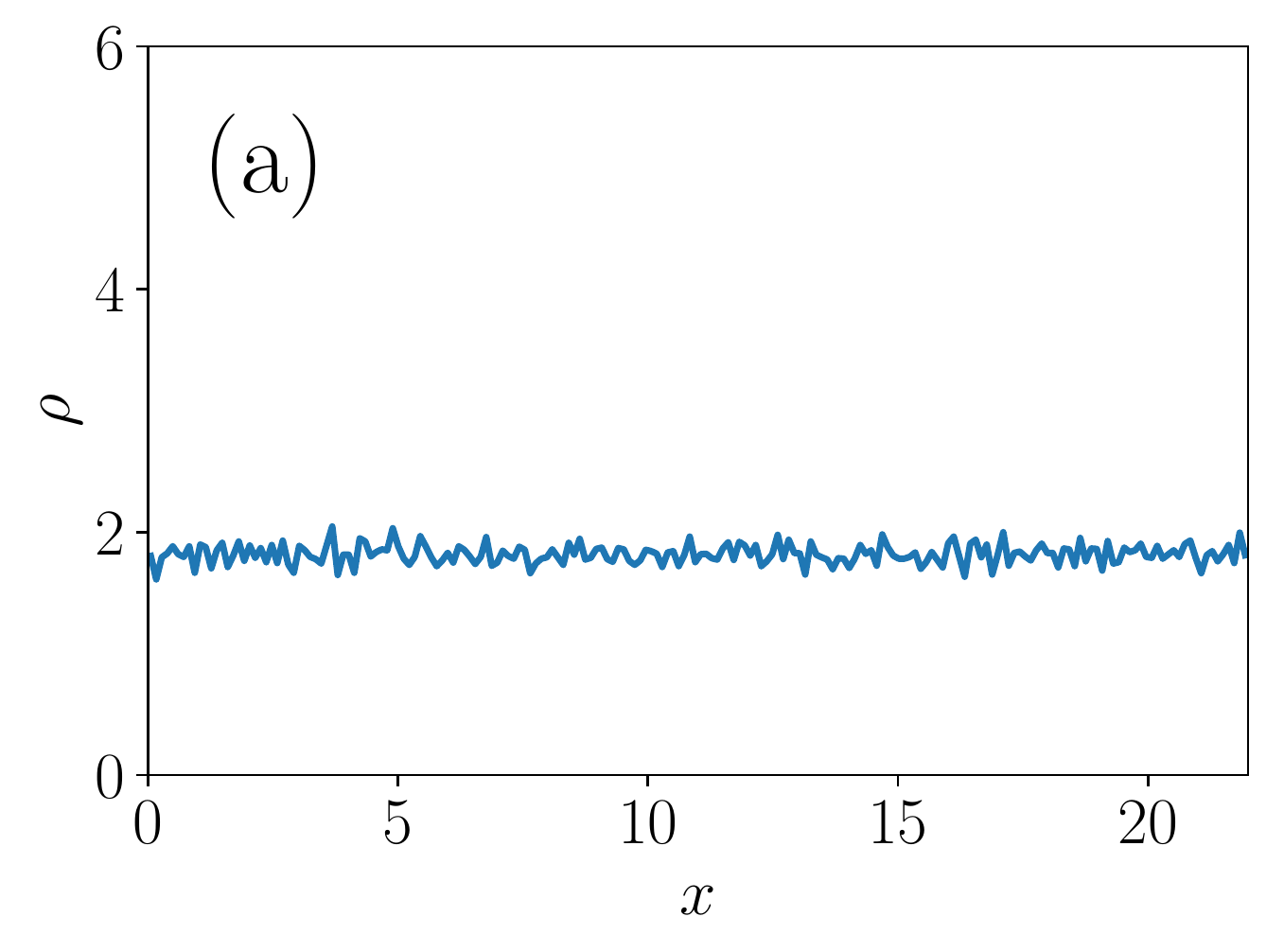}
\end{subfigure}
~
\begin{subfigure}[b]{0.31\textwidth}
\includegraphics[width=1\textwidth]{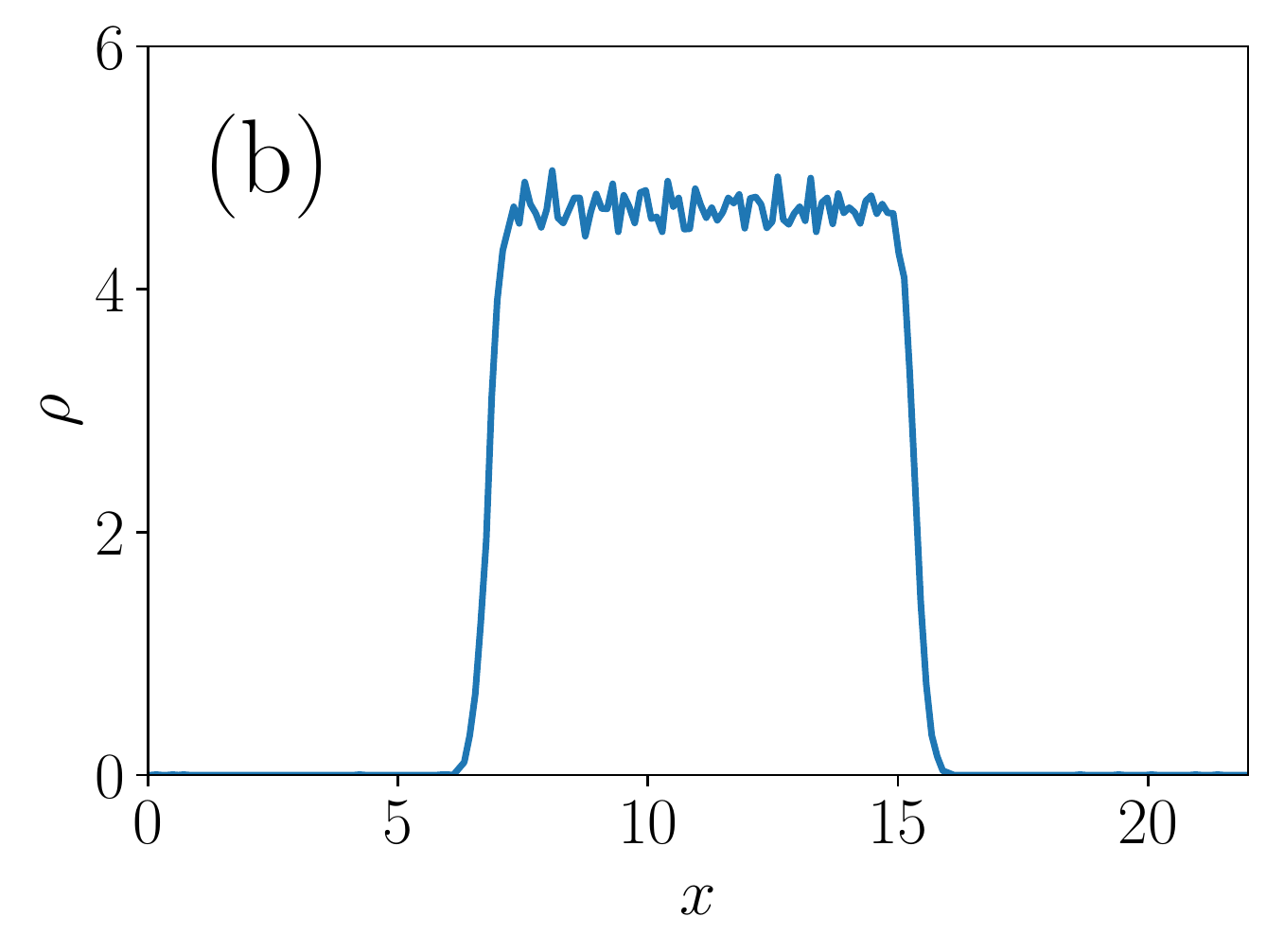}
\end{subfigure}
~
\begin{subfigure}[b]{0.31\textwidth}
\includegraphics[width=1\textwidth]{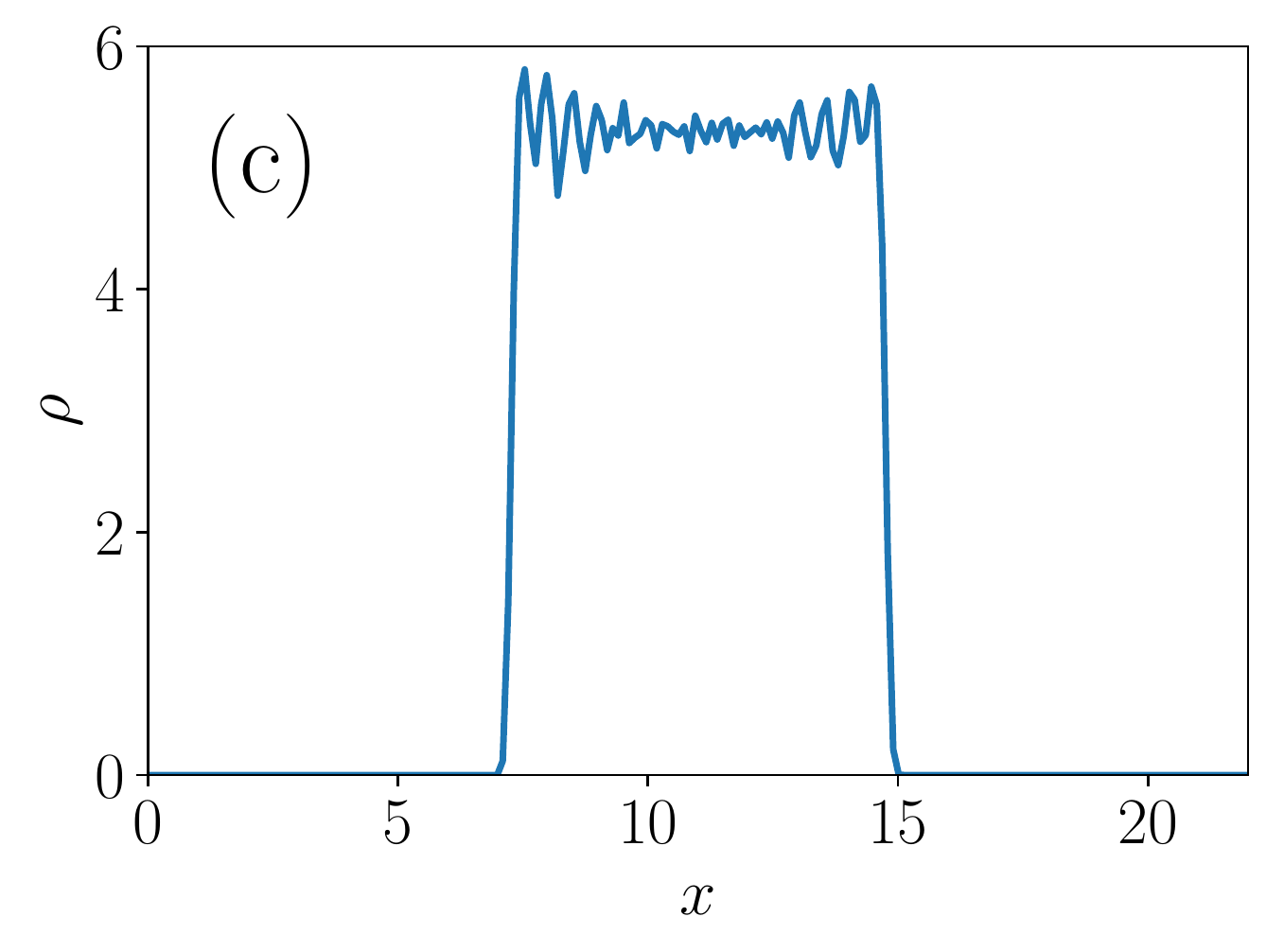}
\end{subfigure}
\caption{Representative density profiles of the MDPD depicting (a) gas phase at $\rd=0.75, A=-5, B=25$, (b) liquid phase at $\rd=0.75, A=-45, B=65$, and (c) solid phase at $\rd=0.75, A=-95, B=100$. From the similarity of (b) and (c) it follows that the solid phase cannot reliably be identified by its density profile.}
\label{fig:profiles}
\end{figure*}

Our first tool in describing the properties of MDPD fluid is density, which arises naturally as a function of the repulsion, attraction and many-body cutoff $\rd$ and not due to the constraints of the simulation cell, as in case of standard DPD. Fig.~\ref{fig:profiles} shows typical density profiles in a cell of size $22\times5\times5$ for $\rd=0.75$ and 0.65. For low values of both $|A|$ and $B$, we observed homogeneously dispersed particles signalling the gas phase. For intermediate values between 0 and 100 there is a liquid phase with a well-defined interface. Finally, the periodic variation of zero and very high density at $\rd=0.65$ indicates a lattice of a solid phase.

To quantify these observations we fitted the density profiles with a symmetrised hyperbolic tangent:
\begin{equation}
\rho_{A, B}(x) = c_1[\tanh(c_2 |x - c_3| + c_4) + 1]/2 + c_5.
\label{eq:tanh}
\end{equation}
where $c_1$ is the excess density of the liquid phase, $c_5$ is the density of the gas phase, $c_3$ and $c_4$ are the centre and the half-width of the liquid droplet respectively, and $c_2$ is the steepness of the interface. The resulting liquid density is then $\rho = c_1 + c_5$.

\begin{figure*}
\centering
\begin{subfigure}[b]{0.24\textwidth}
\includegraphics[width=1\textwidth]{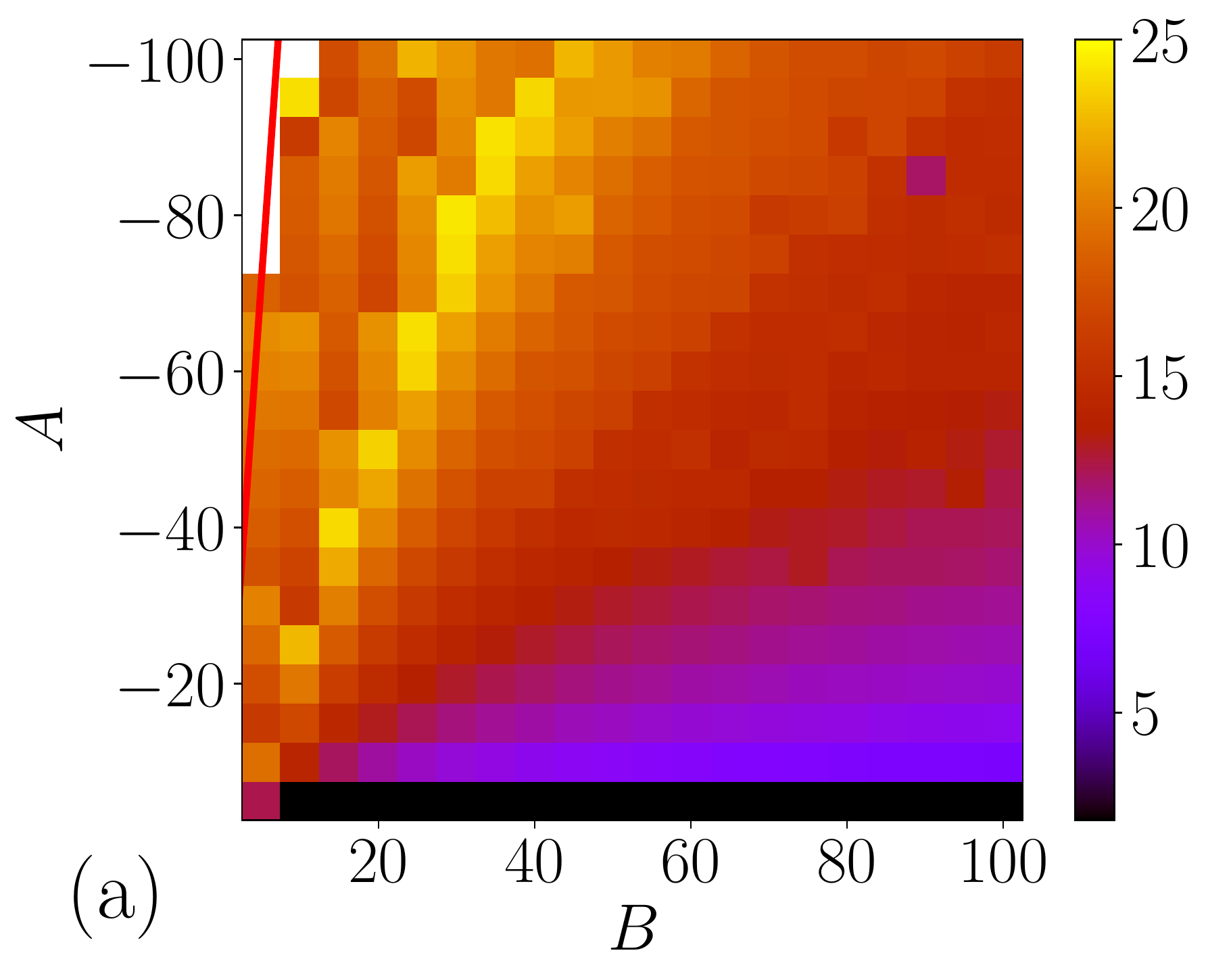}
\end{subfigure}
\hspace{-1mm}
\begin{subfigure}[b]{0.24\textwidth}
\includegraphics[width=1\textwidth]{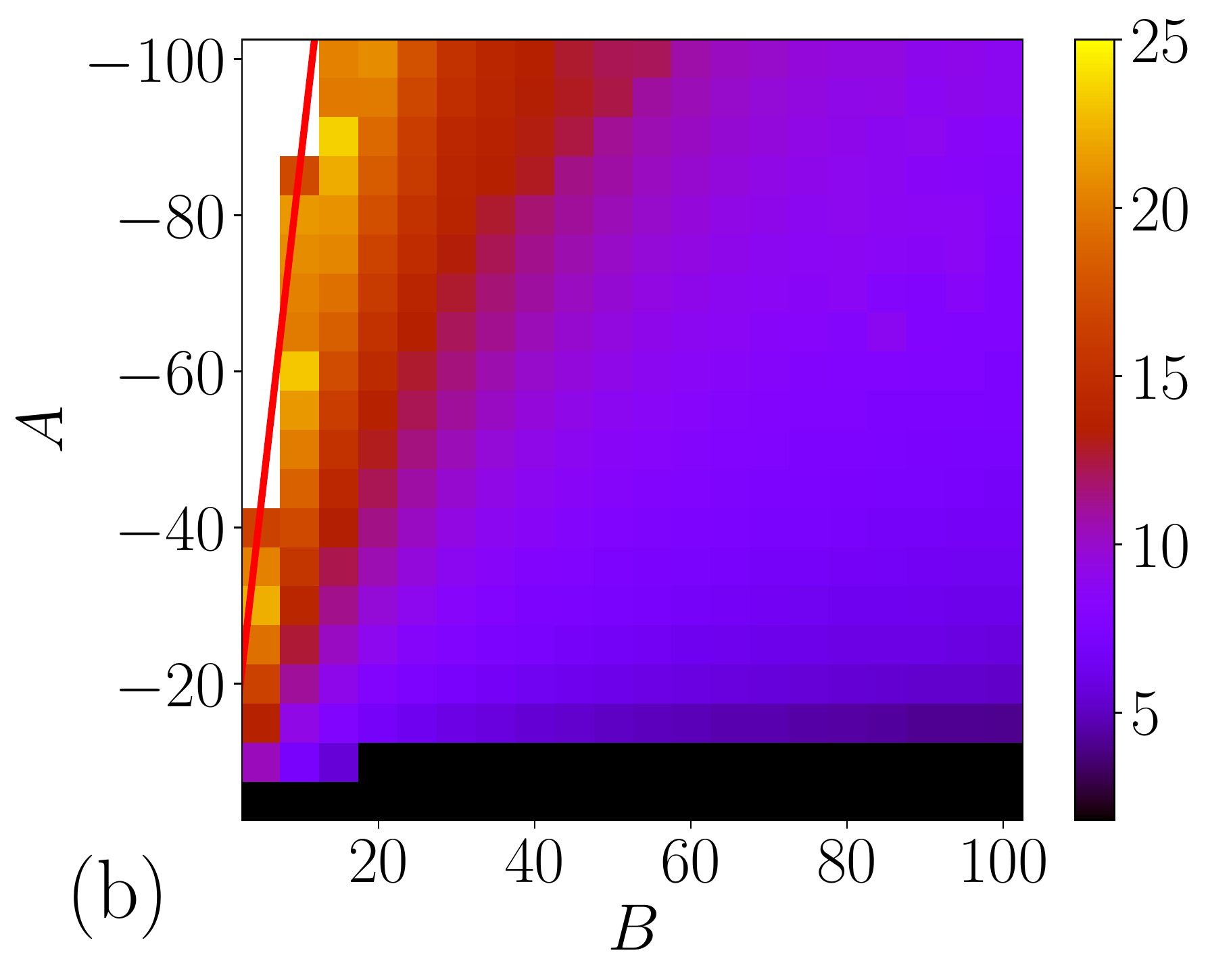}
\end{subfigure}
\hspace{-1mm}
\begin{subfigure}[b]{0.24\textwidth}
\includegraphics[width=1\textwidth]{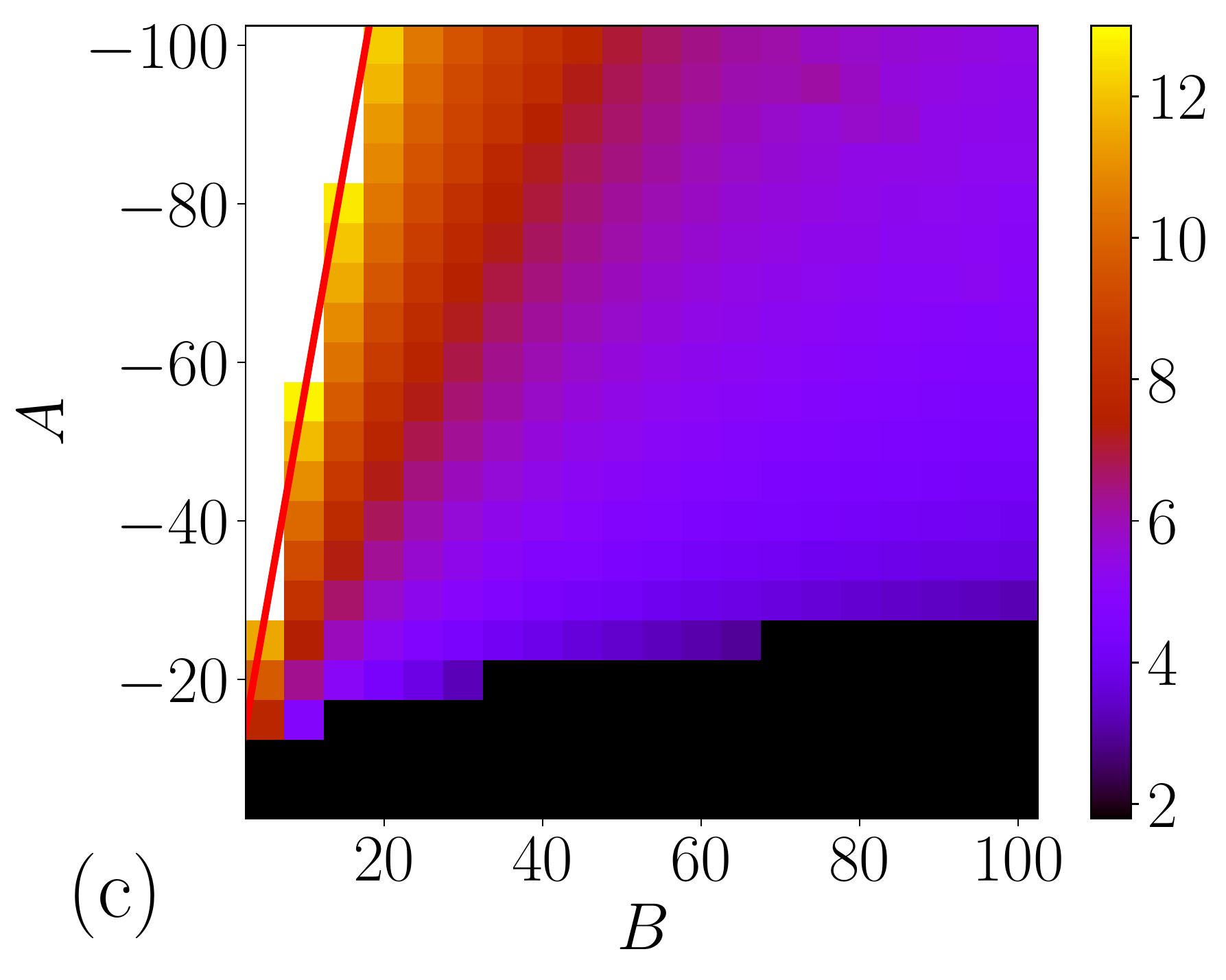}
\end{subfigure}
\hspace{-1mm}
\begin{subfigure}[b]{0.24\textwidth}
\includegraphics[width=1\textwidth]{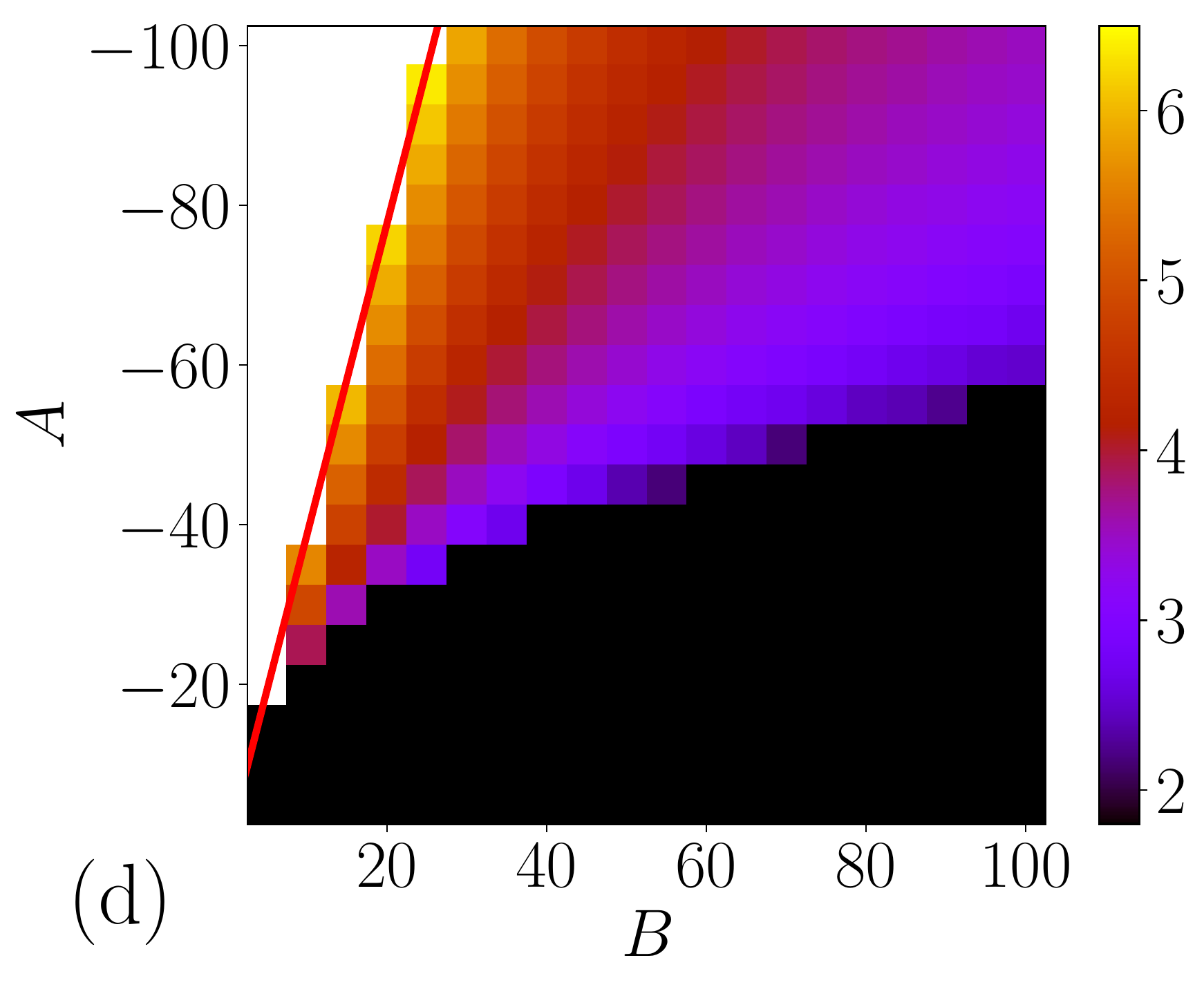}
\end{subfigure}
\caption{(Color online) Density heat maps for (a) $\rd=0.55$, (b) $\rd=0.65$, (c) $\rd=0.75$, and (d) $\rd=0.85$. Dark regions at low values of $|A|$ show the gas phase, and yellow regions of high density shown at the top left corner reveal the no-go region with attractive force at zero interparticle distance.}
\label{fig:rho}
\end{figure*}

Fig.~\ref{fig:rho} shows heat maps of the computed densities $\rho = c_1 + c_5$, with each subfigure representing a specific many-body cutoff. At $\rd=0.85$, the gas phase (dark blue colour) occupies almost one half of the phase diagram, indicating that at higher values of $\rd$ there would be no space for the liquid phase within a reasonable range of repulsions and attractions. On the other hand, at low values of $\rd$, such as 0.55, the gas phase is limited to very low values of $|A|$, and most of the region is occupied by the solid phase, as will be confirmed by self-diffusivity measurements in Section~\ref{sec:diff}.

We now determine how the liquid or solid density vary with the force field parameters. For simplicity, we perform this fitting separately for each value of $\rd$. In principle it is possible to obtain such dependence by analytically finding the roots of the EOS at zero pressure. However, our attempt to solve Warren's EOS (eq.~\eqref{eq:warren}) analytically resulted in an expression that was too long and intractable for further use. Our aim is instead to produce a density function which is more empirical but at the same time more practical for subsequent applications. This can be achieved using only a few fitting parameters and simple polynomial, power law or exponential functions.

Visually observing the cuts through the phase diagram and exploring several candidate functions we arrived at a simple three-parameter fit suitable for all considered many-body cutoffs:
\begin{equation}
\rho(A,B) = d_1 + d_2 (-A) B^{d_3}
\label{eq:rho_fit}
\end{equation}
with fitting coefficients $d_i, i\in\{1,2,3\}$. Their values and associated errors are shown in Table~\ref{tbl:rho_coeffs}. We did not fit the lowest explored value of the cutoff $\rd=0.55$ due to its very small liquid phase, but in principle this can be done as well as for any other cutoff. A more detailed analysis, including the model selection, is provided in the appendix.

\begin{table}
\centering
\begin{ruledtabular}
\begin{tabular}{l|ccc}
$\rd$ & $d_1$ & $d_2$ & $d_3$\\\hline
0.65 & 5.01$\pm$0.03 & 2.11$\pm$0.05 & $-0.870\pm$0.01 \\
0.75 & 3.01$\pm$0.03 & 1.21$\pm$0.03 & $-0.856\pm$0.01 \\
0.85 & 1.50$\pm$0.05 & 0.60$\pm$0.02 & $-0.756\pm$0.01 \\
\end{tabular}
\end{ruledtabular}
\caption{Fitting coefficients for liquid and solid density (eq.~\eqref{eq:rho_fit}) as a function of $A$, $B$, and $\rd$.}
\label{tbl:rho_coeffs}
\end{table}

\subsection{Self-diffusivity}
\label{sec:diff}
To reliably identify the boundary between solid and liquid phase for $\rd\in\{0.65,0,75,0.85\}$, we investigate the dynamic properties of MDPD. The self-diffusivity of an unknown material is an important signature differentiating between liquid, solid and gas phases. We expect this quantity to be negligible in solids, while in pure liquids or gases it should follow the Einstein regime marked by the linear dependence of the mean-square displacement on time.

We measured the self-diffusion coefficient for every configuration via the mean-square displacement (MSD):
\begin{equation}
D=\lim_{t\rightarrow\infty} \frac{\langle |\vc r(t) - \vc r(0)|^2\rangle}{6 t}
\end{equation}
where the average $\langle.\rangle$ is over all the particles.

\begin{figure*}
\centering
\begin{subfigure}[b]{0.32\textwidth}
\includegraphics[width=1\textwidth]{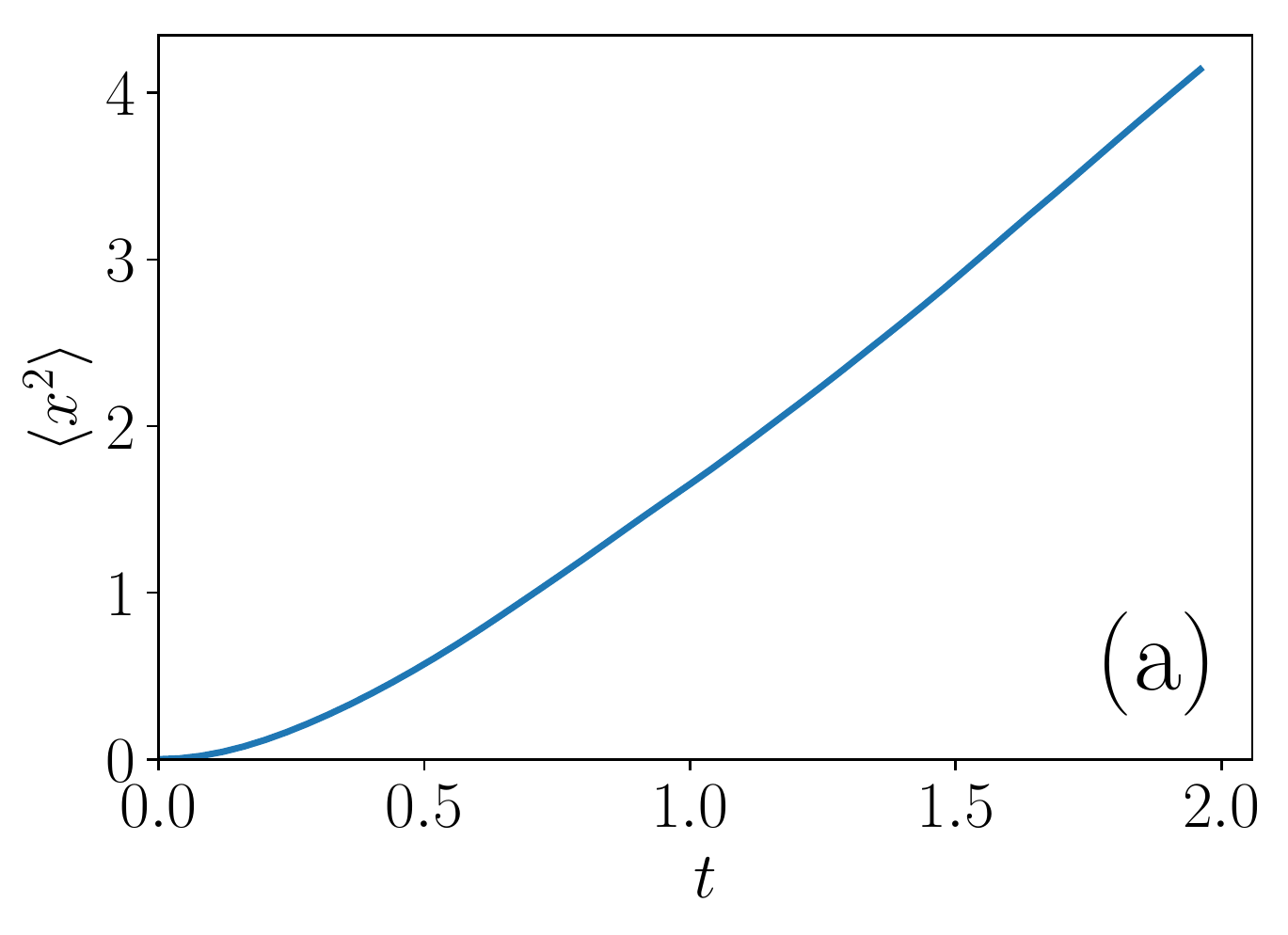}
\end{subfigure}
\hspace{-2mm}
\centering
\begin{subfigure}[b]{0.33\textwidth}
\includegraphics[width=1\textwidth]{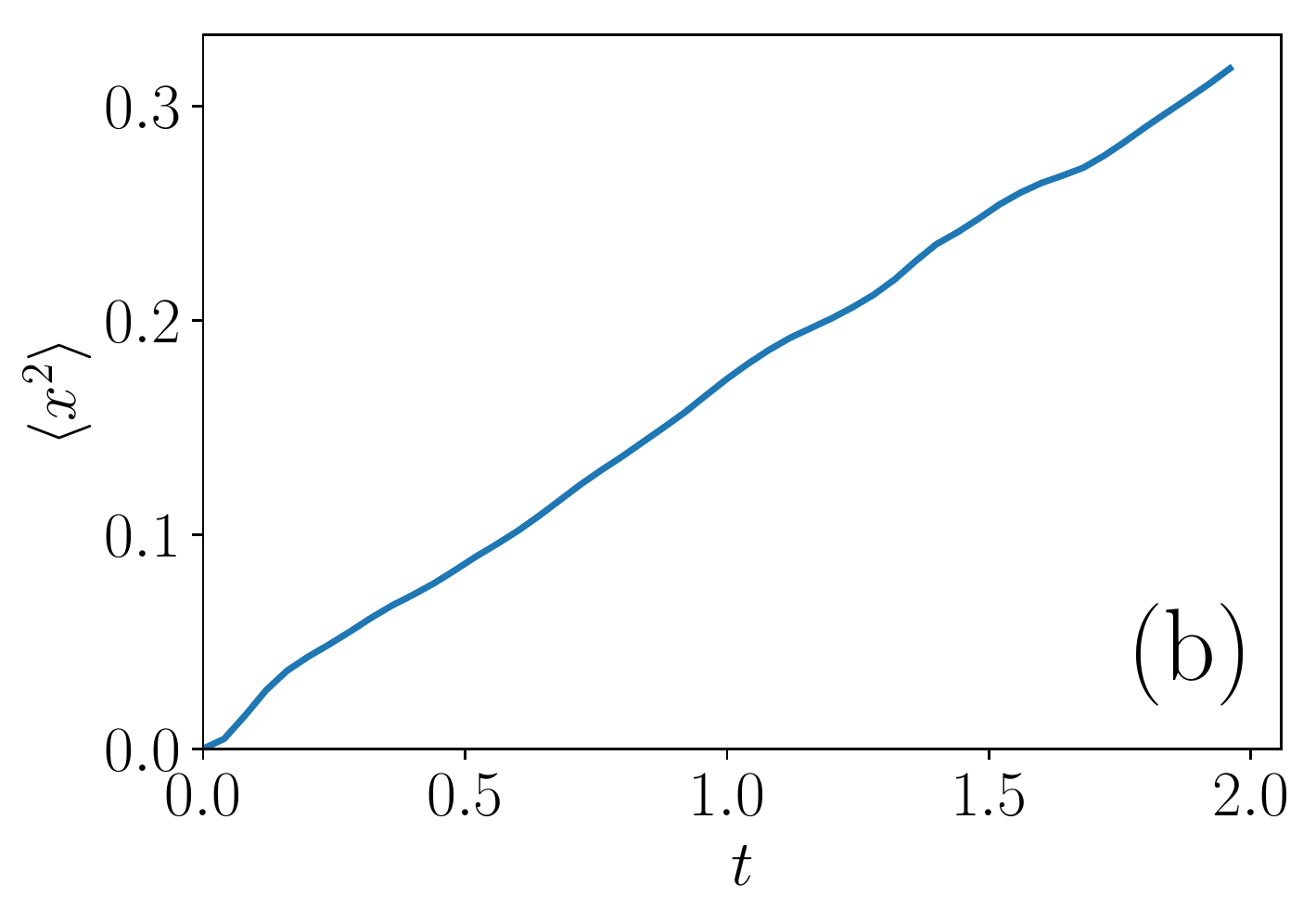}
\end{subfigure}
\hspace{-2mm}
\centering
\begin{subfigure}[b]{0.34\textwidth}
\includegraphics[width=1\textwidth]{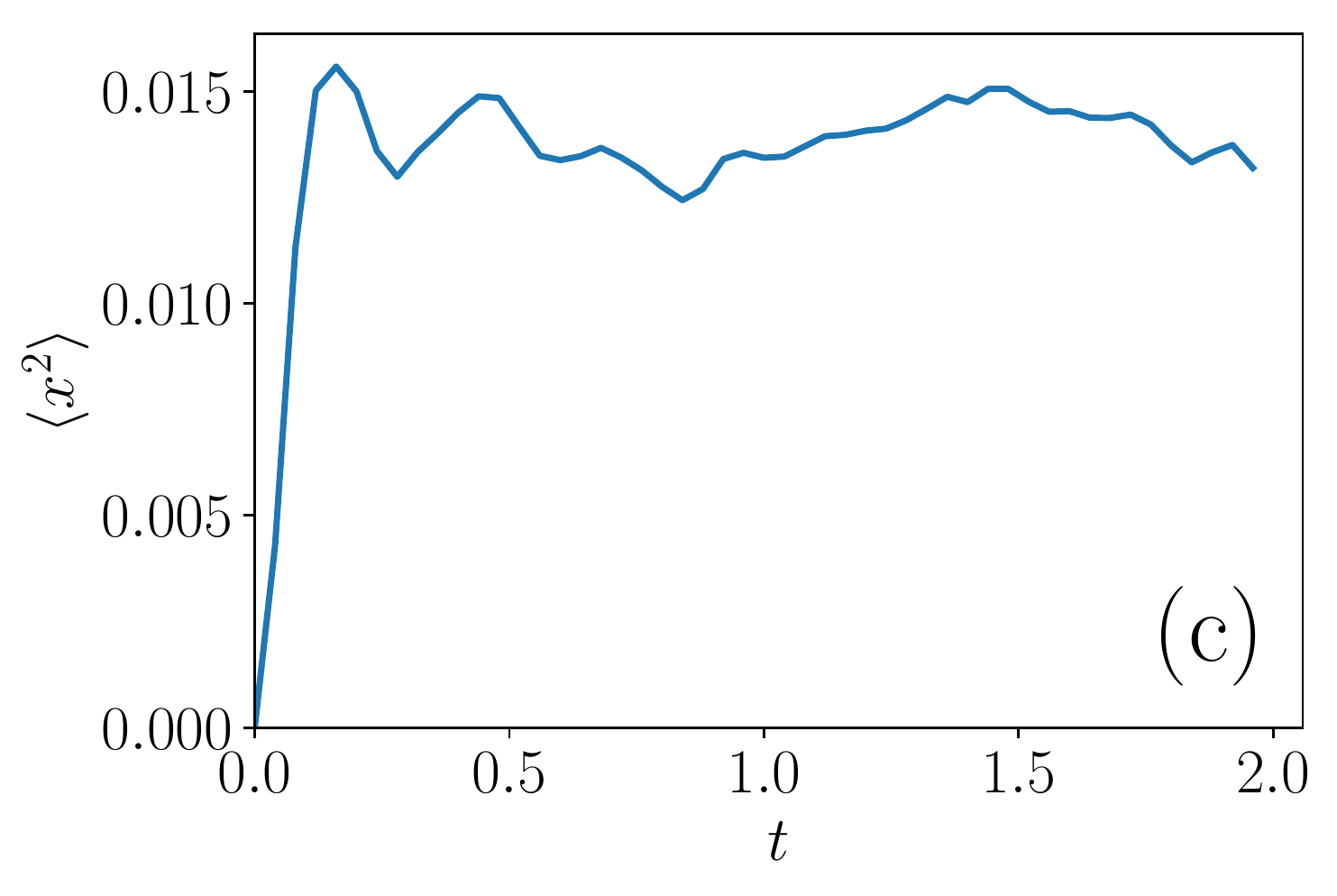}
\end{subfigure}
\caption{Mean-square displacements for the representative density profiles observed in many-body DPD, depicting typical behaviour of (a) gas phase at $\rd=0.75, A=-5, B=25$, (b) liquid phase at $\rd=0.75, A=-45, B=65$, and (c) solid phase at $\rd=0.75, A=-95, B=100$.}
\label{fig:msd}
\end{figure*}

Typical MSDs are shown on Figs~\ref{fig:msd}. The scale on the y-axis demonstrates a clear difference between solids, liquids and gases. The solid phase poses a limit to the beads in how far they can diffuse. The liquid phase allows only the linear regime, whereas the gas phase contains a polynomial transient response and then gradually becomes linear.

\begin{figure*}
\centering
\begin{subfigure}[b]{0.25\textwidth}
\includegraphics[width=1\textwidth]{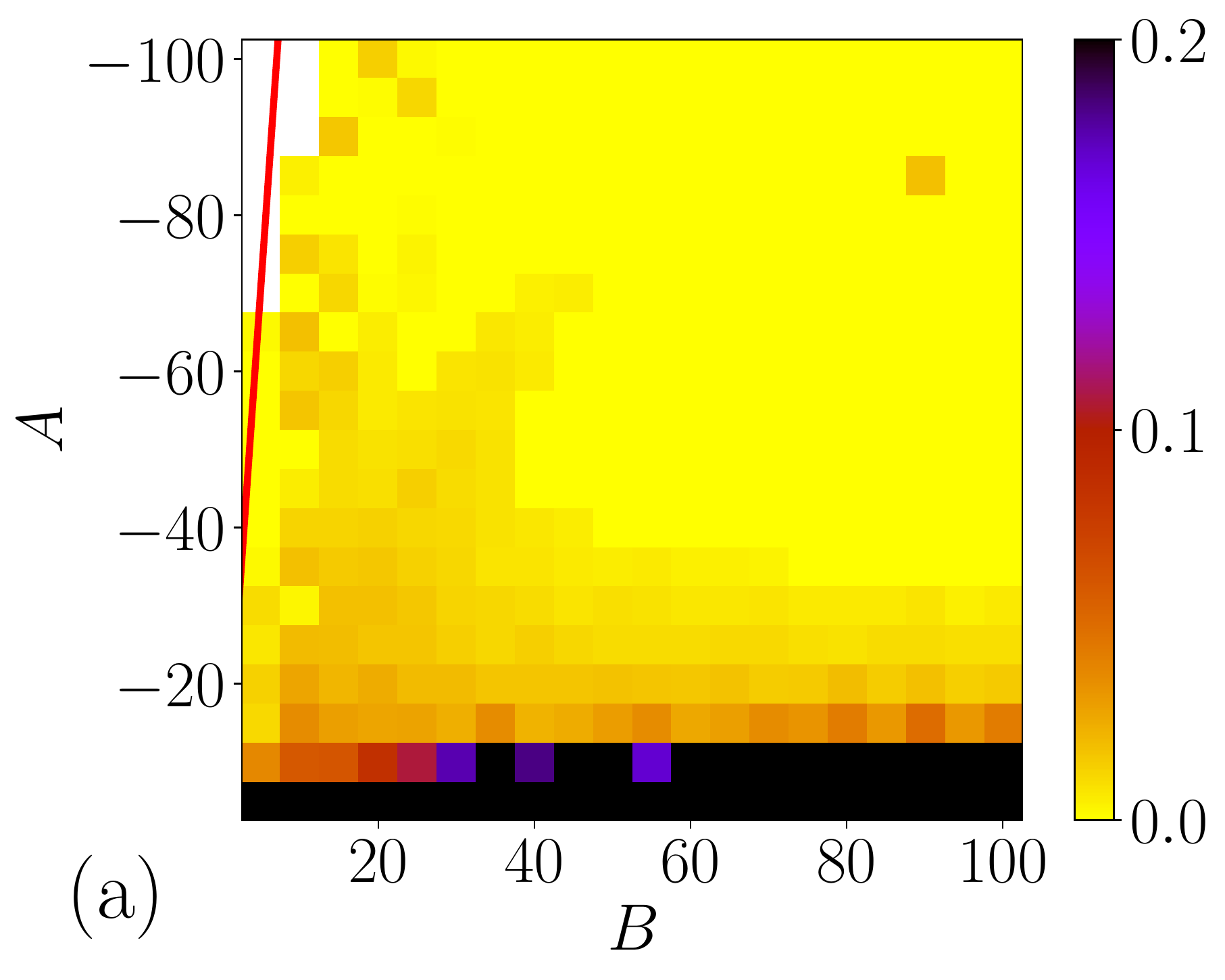}
\end{subfigure}
\hspace{-2mm}
\begin{subfigure}[b]{0.25\textwidth}
\includegraphics[width=1\textwidth]{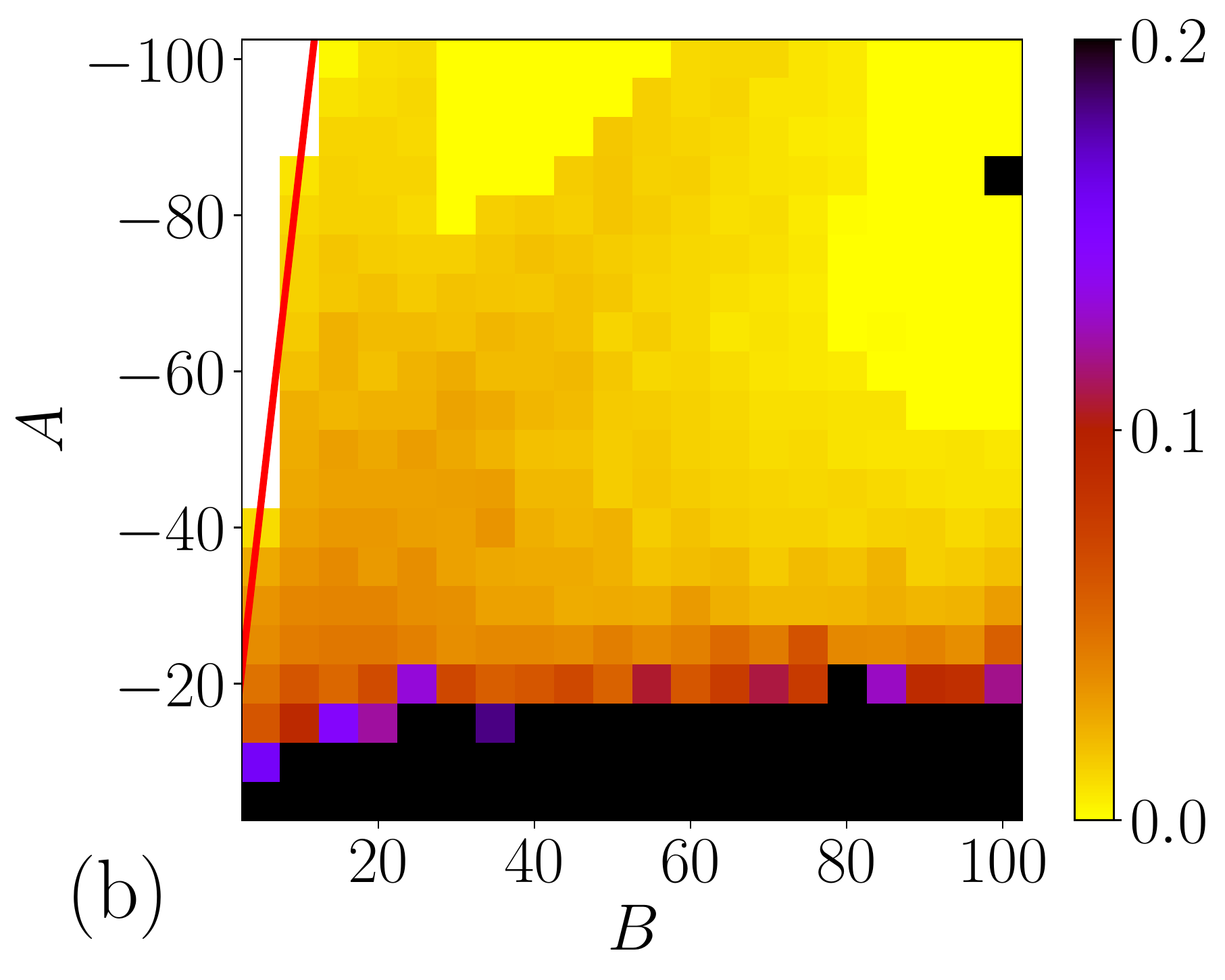}
\end{subfigure}
\hspace{-2mm}
\begin{subfigure}[b]{0.25\textwidth}
\includegraphics[width=1\textwidth]{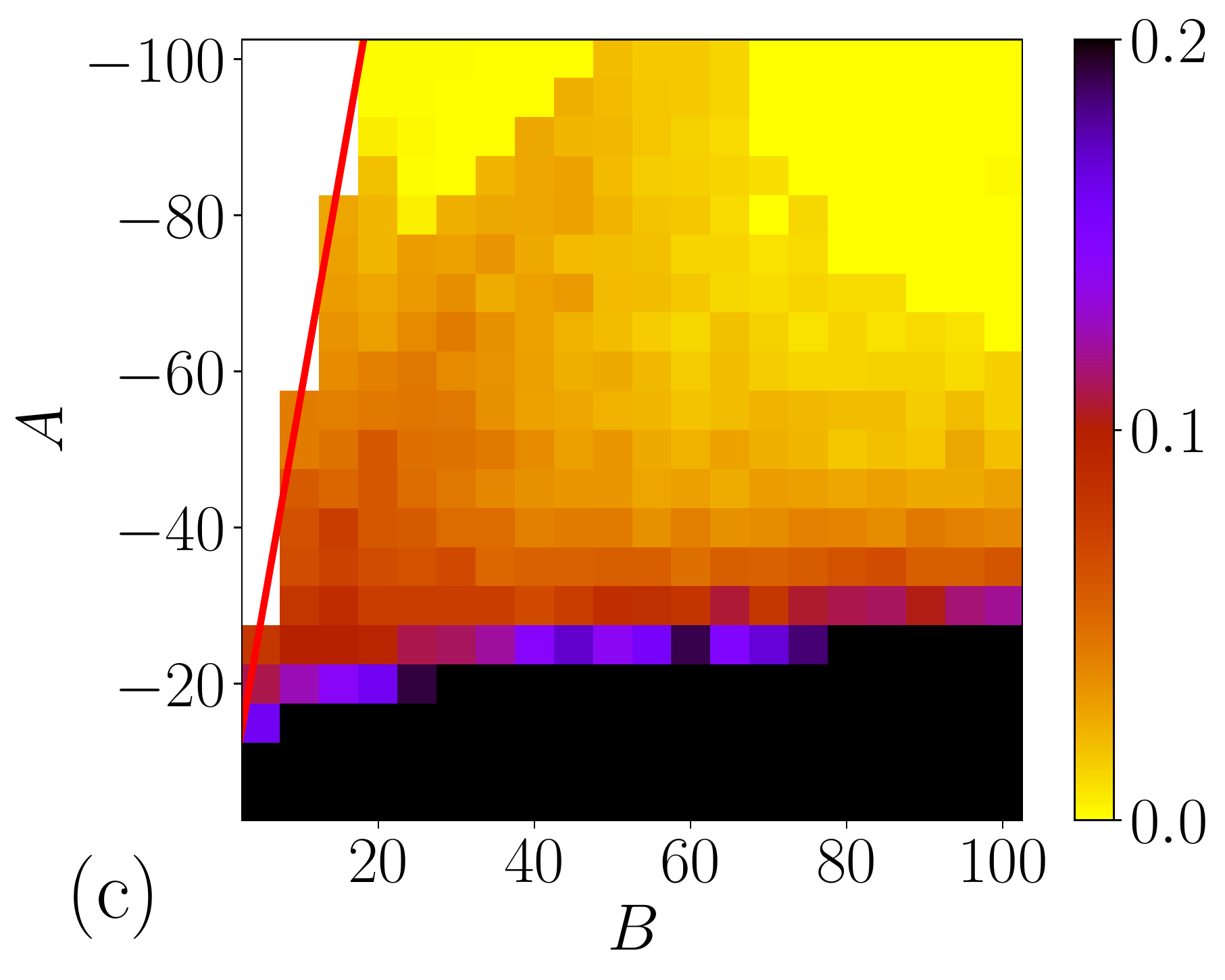}
\end{subfigure}
\hspace{-2mm}
\begin{subfigure}[b]{0.25\textwidth}
\includegraphics[width=1\textwidth]{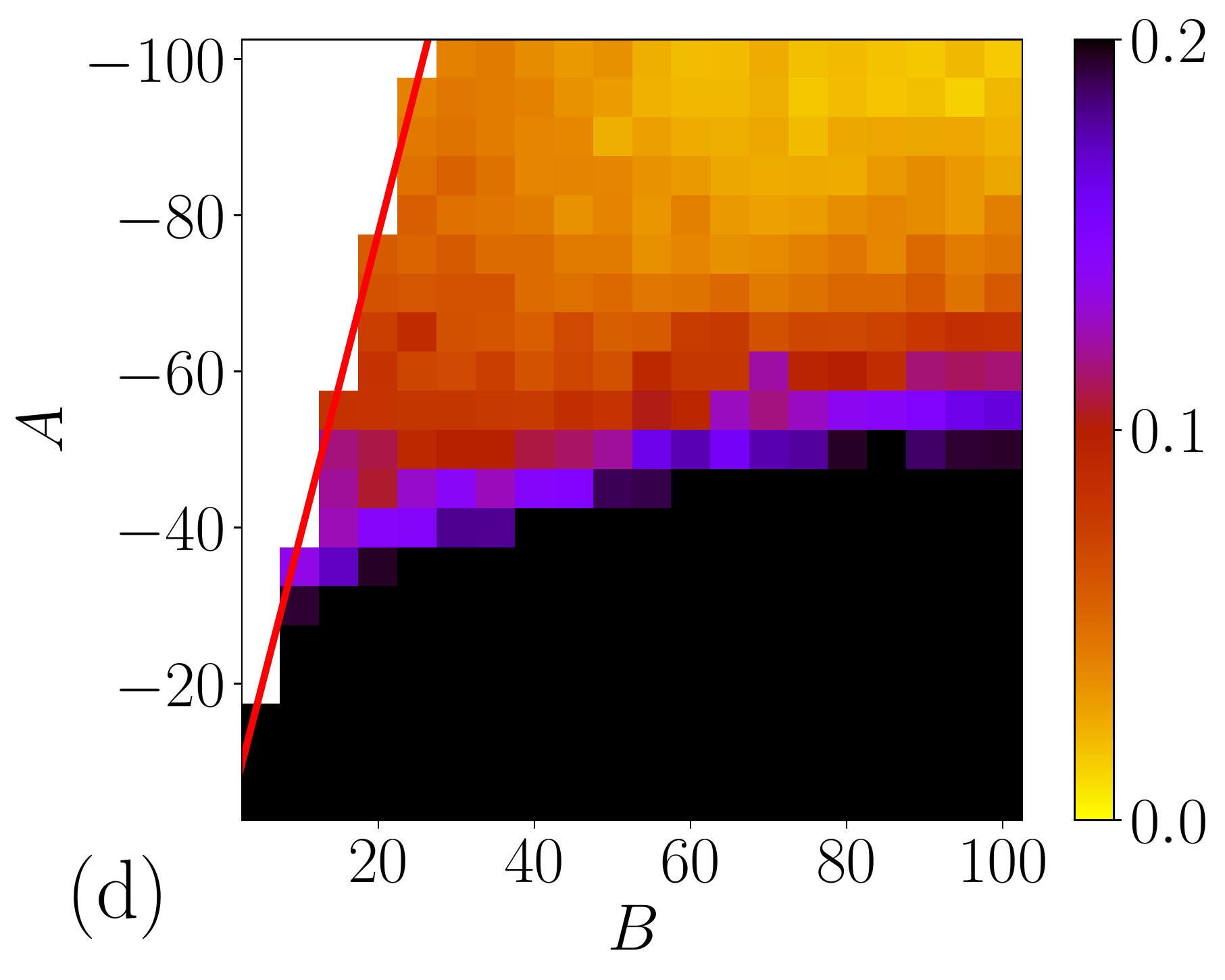}
\end{subfigure}
\caption{(Color online) Self-diffusivity heat maps for (a) $\rd=0.55$, (b) $\rd=0.65$, (c) $\rd=0.75$, and (d) $\rd=0.85$. Yellow regions at the top reveal the solid phase; dark regions at the bottom show the gas phase.}
\label{fig:diff}
\end{figure*}

Plotting all the self-diffusivities in a heat map (Figs~\ref{fig:diff}) enables us to distinguish the different phases. Dark blue regions corroborate the existence of the solid phase, whereas the yellow regions show the gas phase. The region in between is liquid.

We also briefly probe the nature of the boundary between the liquid and the gas phase. Having chosen several values of the repulsion $B$ and finely varying the attraction $A$, we monitored the points at which the denser liquid droplet started to rise from a homogeneous gas. For $B>20$, the liquid-gas boundary is well captured by a line:
$A_{\rm{lg}}=\omega_1 B + \omega_2$. For example, at $\rd=0.75$, the fitting constants are $(\omega_1,\omega_2)=(-0.13,15.3)$.

\subsection{Lattice of the solid phase}
Having located the whereabouts of the solid phase in the phase diagram via the self-diffusion coefficient, we now determine its lattice. There are in fact two lattice types, implying two different phases. Starting with $\rd=0.75$, we observe the first type occurring at large values of both repulsion and attraction, around $(A,B)=(-100,\,100)$. The density of this configuration is $\rho\approx 5$. Another phase, which is formed at high repulsions $A<-80$ and intermediate attractions $B = 30$--50, is more closely packed, with a typical density of $\rho\approx 8$ at $\rd=0.75$. 

\begin{figure}
\centering
\begin{subfigure}[b]{0.45\textwidth}
\includegraphics[width=1\textwidth]{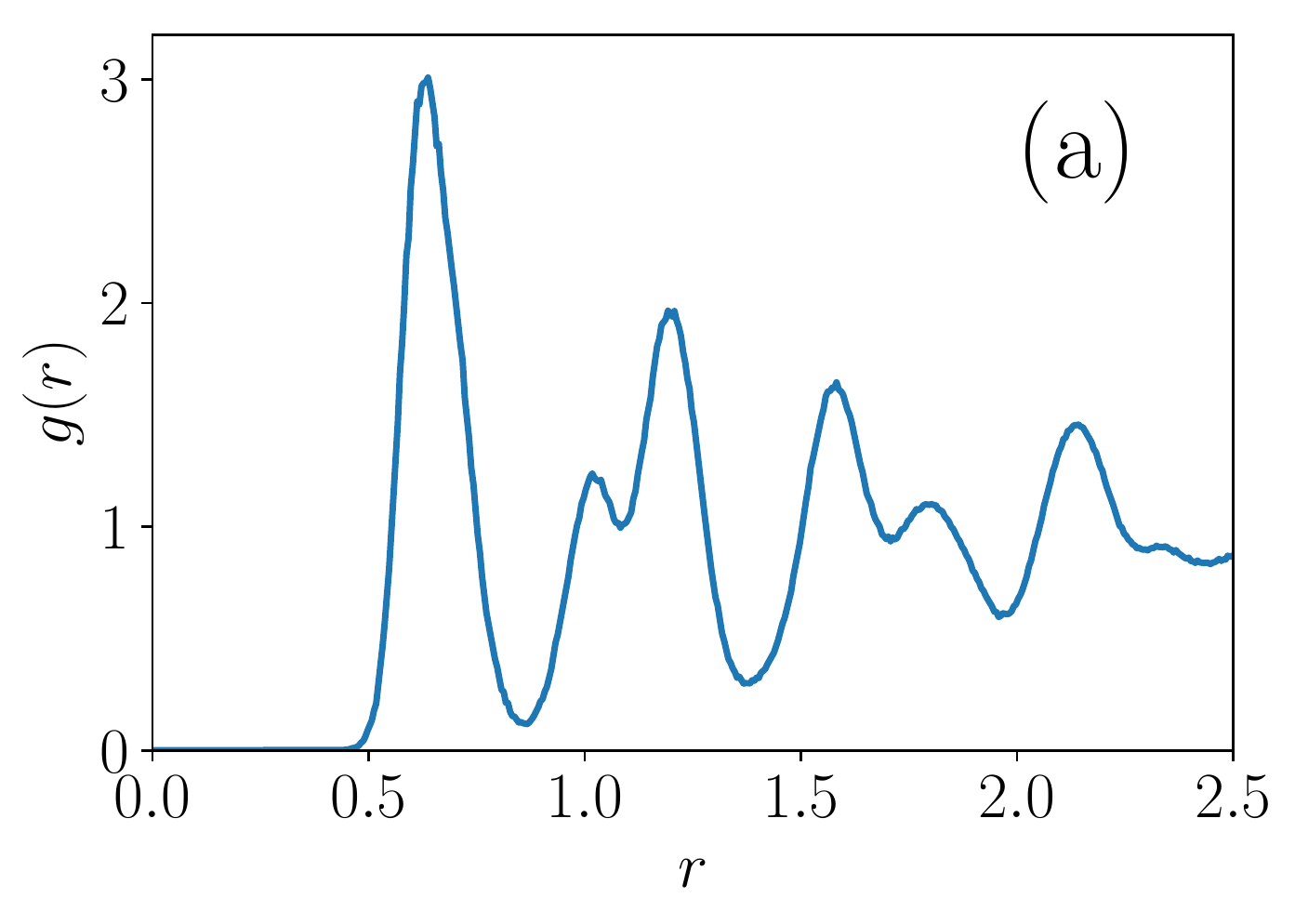}
\end{subfigure}
\\
\begin{subfigure}[b]{0.2\textwidth}
\includegraphics[width=1\textwidth]{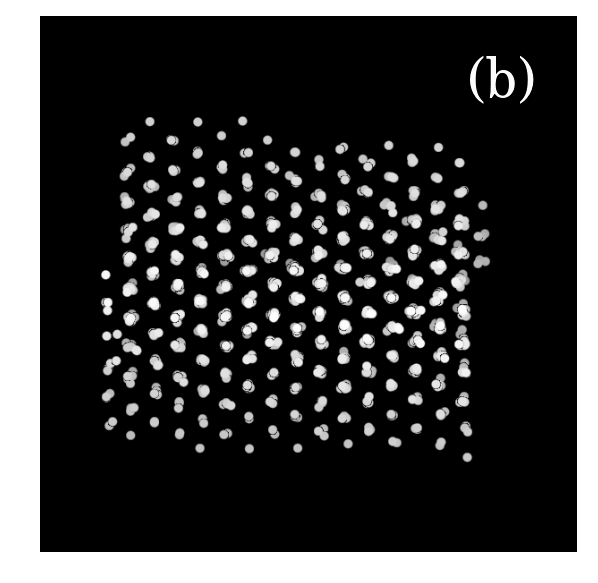}
\end{subfigure}
~
\begin{subfigure}[b]{0.2\textwidth}
\includegraphics[width=1\textwidth]{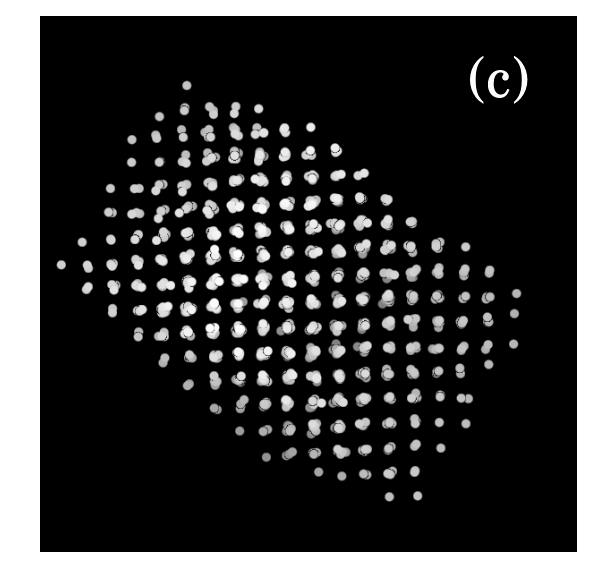}
\end{subfigure}
\caption{(a) Radial distribution function of the bcc phase for parameters $\rd=0.75, A=-100, B=100$. (b) and (c) show lattice visualisations of the bcc phase.}
\label{fig:lattice}
\end{figure}

To identify these phases, we plotted the radial distribution functions (RDF) and compared them with a set of RDFs of several Bravais lattices smeared by temperature fluctuations. The first phase was identified with the body-centred cubic (bcc) lattice (Fig.~\ref{fig:lattice}), and the second one with the hexagonal (hex) lattice with an interlayer distance lower than the in-plane lattice constant.

As another verification, we computed the coordination numbers (CN) for all the solid configurations $(A,B,\rd)$, which we chose by their self-diffusivity. CN is defined as the number of nearest neighbours of a particle, which can be computed by integrating the RDF $g(r)$:
\begin{equation}
z(r) = \rho \int_0^r g(r') \, 4\pi r'^2\, dr'.
\end{equation}

\begin{figure*}
\centering
\begin{subfigure}[b]{0.30\textwidth}
\includegraphics[width=1\textwidth]{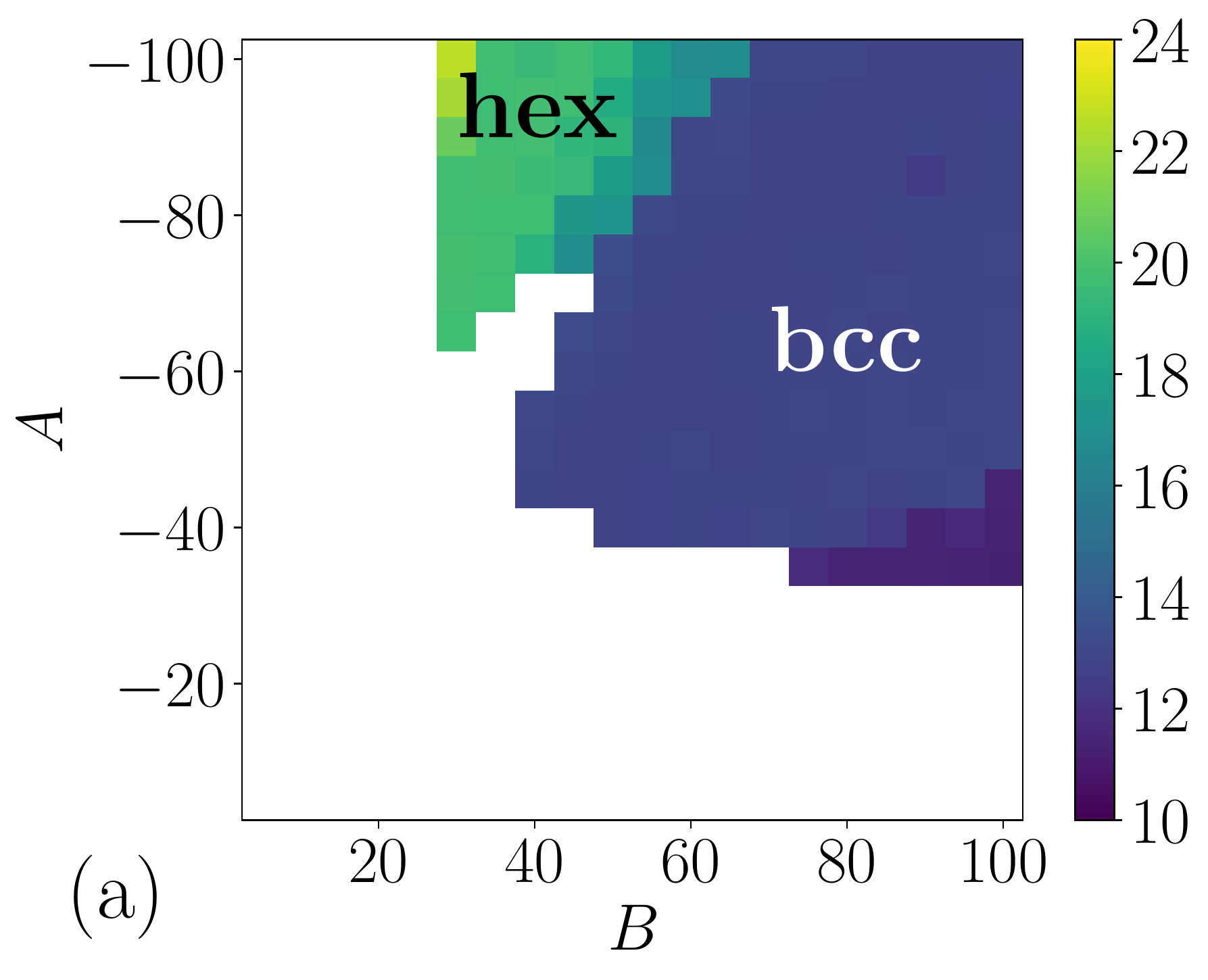}
\end{subfigure}
~
\begin{subfigure}[b]{0.30\textwidth}
\includegraphics[width=1\textwidth]{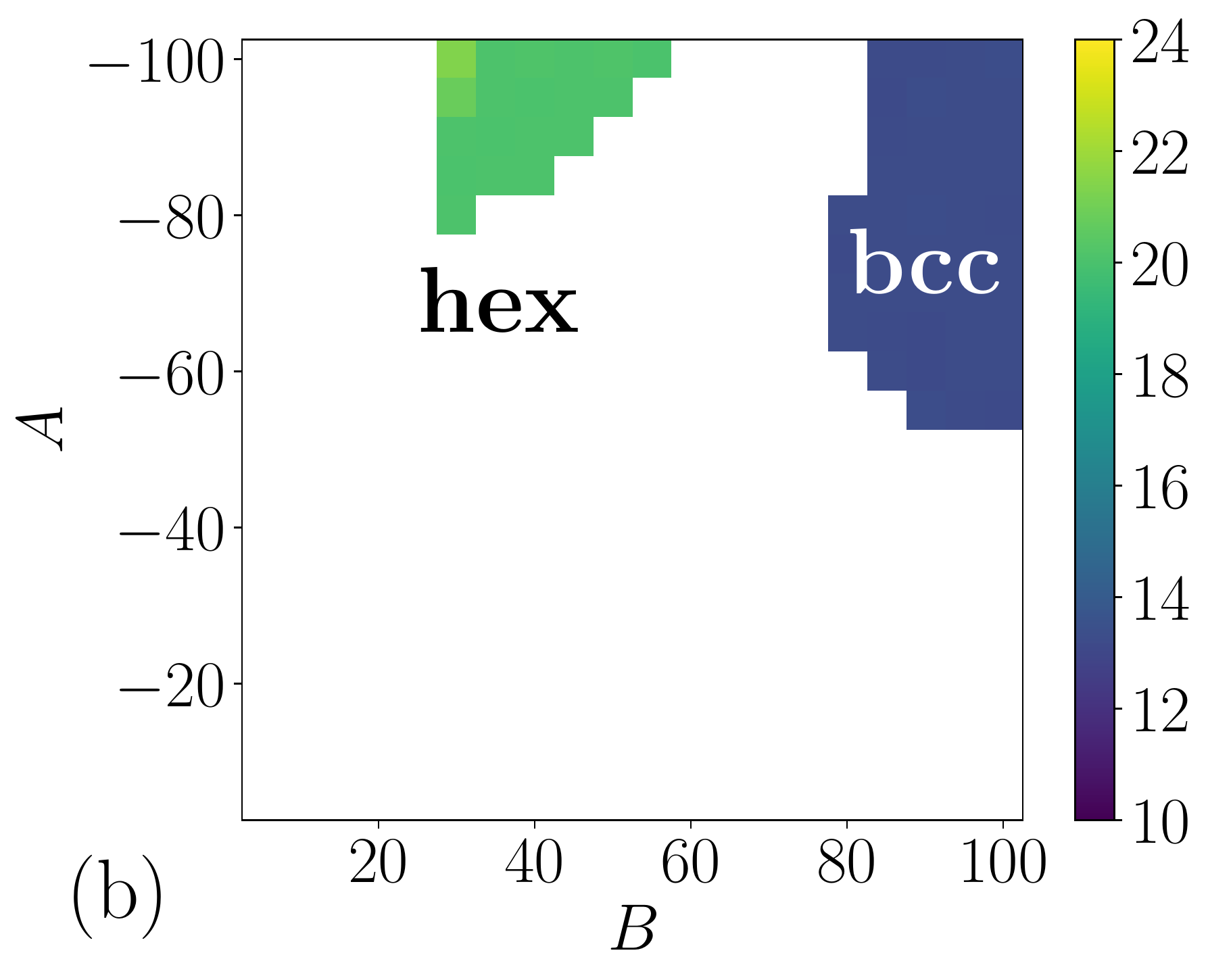}
\end{subfigure}
~
\begin{subfigure}[b]{0.30\textwidth}
\includegraphics[width=1\textwidth]{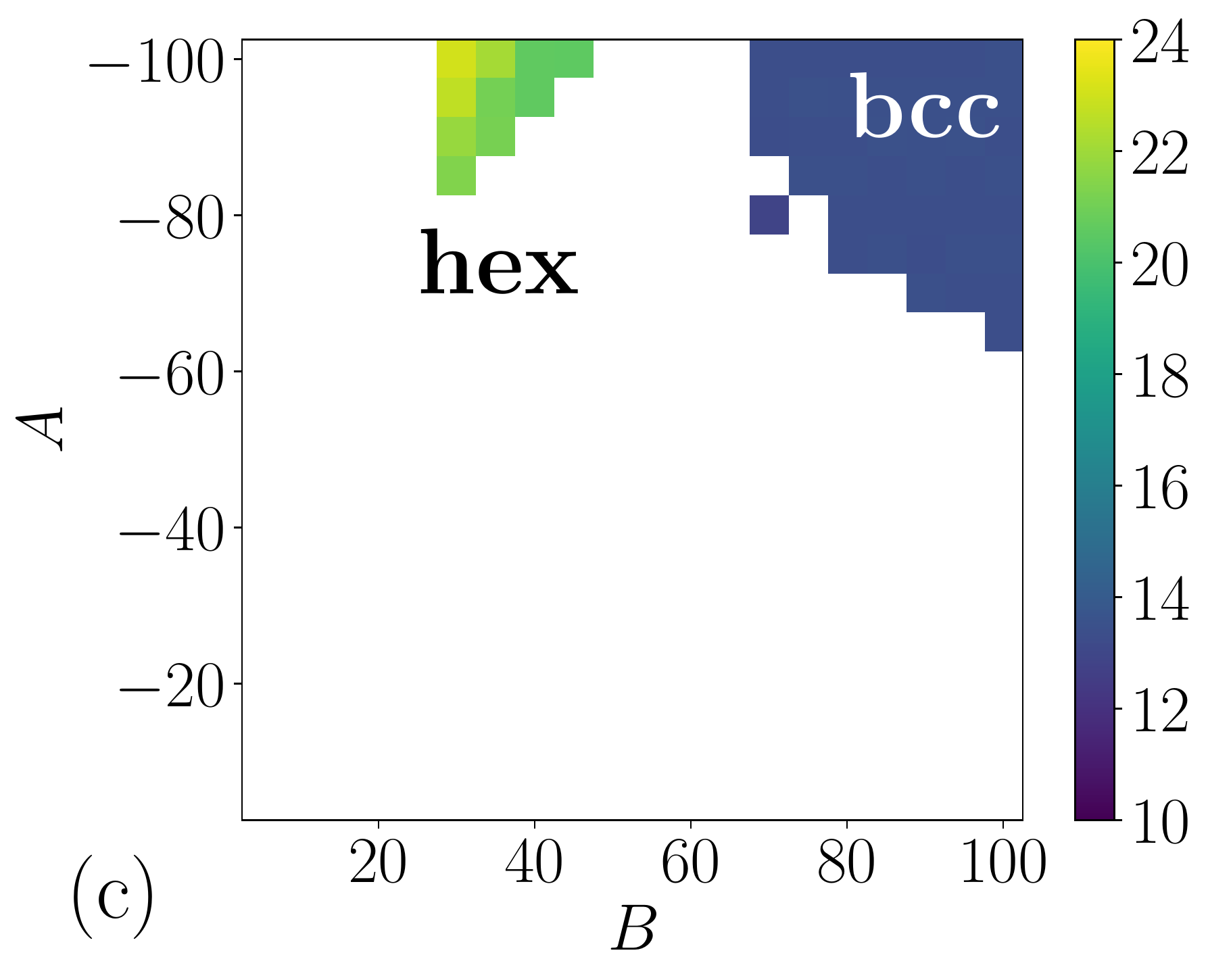}
\end{subfigure}
\caption{(Color online) Heat maps of the coordination numbers for various many-body cutoffs (a) $\rd=0.55$, (b) $\rd=0.65$, (c) $\rd=0.75$, which contain a solid phase, with the lattice denoting a specific phase type.}
\label{fig:cn}
\end{figure*}

In a lattice, neighbouring particles reside in so-called coordination shells, which give rise to local maxima in the RDF. Separating the adjoining coordination shells can be realised by identifying the plateaus in the CN as a function of the distance, \emph{i.e.} the minima in the first derivative of $z(r)$. Fig.~\ref{fig:cn} unambiguously shows that all the solid configurations $(A,\,B,\,\rd)$ indeed fall into two groups: the bcc phase with a plateau value of $z \approx 14$, which includes first two coordination shells, and the hex phase with a first plateau $z=2$, which captures out-of-plane vertically aligned atoms, followed by $z\approx 20$, which comprises two hexagons above and below and one in the plane of any particle.

From Fig.~\ref{fig:cn} it is also clear that the solid phase occupies a major part of the phase diagram at $\rd=0.55$, rendering the usefulness of this value of this many-body cutoff rather limited for simulations of liquids. On the opposite end, at $\rd=0.85$, the solid phase is non-existent within the explored range of repulsions and attractions. From these observations it follows that most practical for simulation of multiphase systems, as well as richest in terms of the number of phenomena to capture, are simulations at $\rd=0.75$, which has already been widely employed in the literature, as well as 0.65.

We further investigated the stability of both phases, performing simulations in multiple orthorhombic simulation cells of varying degree of asymmetry, between $16\times 4\times 4$ up to the cubic shape, $16\times 16\times 16$, and for a range of densities. For the bcc phase, we took the configurations $(A,\,B,\,\rd)=(-100,\,100,\,0.75)$, at which the equilibrium density was $\rho_{\rm{bcc}} \approx 5.5$. When setting the initial density to around 5.5, the randomly initialised particles indeed formed a bcc lattice for every cell box shape, implying a stable minimum.

To reproduce the hex phase, we chose the configuration $(-100,\,40,\,0.75)$ leading to the equilibrium density $\rho_{\rm{hex}}=8.5$. Starting again from randomly initialised positions, the hex phase formed only when the initial density was set below $\rho_{\rm{hex}}$, and only in the more asymmetric cells. This suggests that the hex phase is stabilised by the negative pressure.

Further investigation by measuring excess chemical potential via the Widom particle insertion method~\cite{Frenkel_book_2002} revealed that the bcc phase is significantly more stable than the hex phase at both $(A,B,\rd)=(-100,\,100,\,0.75)$ and $(-100,\,40,\,0.75)$. We can hence safely conclude that the hex phase is metastable and cannot be considered as a true bulk phase of the MDPD force field.

Finally, to estimate the stress-strain relation of the solid phase, we put an already solid cuboid into a larger simulation cell. After a short simulation period, its shape became spherical. Hence, the true stress-strain relationship of the solid phase cannot be captured by MDPD~\footnote{We thank an unknown reviewer for inspiring this analysis.}.

\subsection{Liquid phase and surface tension}
We now return to the examination of the liquid phase by excluding solid and gas regions. We computed the surface tension for each configuration as follows~\cite{Tolman_JCP_1948, Kirkwood_JCP_1949}:
\begin{equation}
\sigma = \frac{L_x}{2} \bigg(\langle p_{xx}\rangle - \frac{\langle p_{yy} \rangle + \langle p_{zz}\rangle}{2} \bigg),
\end{equation}
where $p_{\beta\beta}$ are the diagonal components of the pressure tensor. As in case of density, we obtain the functional dependence of the surface tension by fitting over the measured points for each many-body cutoff $\rd$. Visual observation of the cuts through the phase diagram at constant $A$ or $B$ and trial of several functions revealed that different many-body cutoffs $\rd$ are best fit by different functions with varying number of parameters. Table~\ref{tbl:sigma_coeffs} summarises these functions and their coefficients. We explain the reasoning for the model selection more fully in the appendix.

\begin{table*}
\centering
\begin{ruledtabular}
\begin{tabular}{c|ll}
$\rd$ & Function & Coefficients \\\hline
0.65 & $(f_1 A^2 + f_2 A + f_3)(B-f_4 + f_5 A)^{f_6}$ & 
$(0.0592, -4.77, -66.8, -1.62, 0.146, -0.665)$\\
0.75 & $(f_1 A^2 + f_2 A)(B + f_3 A)^{f_4}$ & 
$(0.0807, 0.526, 0.0659, -0.849)$\\
0.85 & $(f_1 A^2 + f_2 A)(B-f_3)^{f_4}$     & 
$(0.0218, 0.591, 7.52, -0.803)$\\
\end{tabular}
\end{ruledtabular}
\caption{Fitting functions and their coefficients for the surface tension dependence on $A$ and $B$.}
\label{tbl:sigma_coeffs}
\end{table*}

\section{The connection to real liquids}
\label{sec:real}
Having described the phase diagram of an MDPD fluid and determined the dependence of density and surface tension on the force field parameters $A,B$, and $\rd$, we now discuss how these findings can be used in parametrising real liquids. In the standard DPD, the simulation of a pure fluid is controlled by one parameter $A>0$, and hence only one physical quantity is needed to bridge the simulation with the experiment. Groot and Warren chose compressibility~\cite{Groot_JCP_1997}, but in principle many other experimental properties could be used.

In developing the parametrisation for MDPD, we first assume that $\rd$ is fixed. There remain two free parameters, repulsion and attraction, and so two physical quantities are needed. Having obtained functional relations for density and surface tension over a wide range of configurations $(A,B,\rd)$, we now understand how the behaviour of the liquid, gas or solid varies with the interparticle potential. Furthermore, compressibility is readily available as a function of density and $(A,B,\rd)$ from the EOS (eq.~\eqref{eq:jamali}):

\begin{multline}
\kappa^{-1} = \rho\frac{\partial{p}}{\partial{\rho}} = 
\rho + 2\alpha A \rho^2 \\
+ 2\alpha B\rd^4 (3\rho^3 - 2 c' \rho^2 + d'\rho) 
- \frac{\alpha B\rd^4}{|A|^{1/2}}2\rho^2,
\label{eq:kappa}
\end{multline}
where $\kBT$ was set to 1.

Starting from the interaction parameters in reduced units, we can verify that the relations for density, surface tension, and compressibility yield meaningful liquid properties. As an example, let us take $(A,B,\rd)=(-40,\,25,\,0.75)$, which were first used by Warren to demonstrate the MDPD capabilities by forming a pendant drop~\cite{Warren_PRE_2003}, and later by Ghoufi and Malfeyt to prove that MDPD is capable of simulating liquid water~\cite{Ghoufi_PRE_2011}. Using the values from Table~\ref{tbl:rho_coeffs} we obtain the density 6.09, which is almost equal to the simulation value 6.08 (also obtained by Arienti~\cite{Arienti_JCP_2011}). Employing the appropriate equation and coefficients from Table~\ref{tbl:sigma_coeffs}, the surface tension is 7.01 in reduced units.

To convert these numbers into experimental values, we need to define the reduced units. Following Groot and Rabone's definition of the units in standard DPD simulations~\cite{Groot_BiophysJ_2001}, these depend on the simulated liquid and are based on the average volume per molecule $V_0$, the number of molecules in a bead (a CG degree) $\Nm$, and the target density $\rho$:
\begin{equation}
\rc = (\rho\Nm V_0)^{1/3}.
\label{eq:rc}
\end{equation}
Having determined $\rho$ from $(A,\,B,\,\rd)$ and taking $\Nm=3$, the length scale $\rc$ is 0.818~nm. The experimental observables are summarised in Table~\ref{tbl:real_liq}. The density in SI units is trivially 997 kg/m$^{-3}$, as this is the value on which the parametrisation was based in the form of the volume per molecule $V_0$.

To convert the compressibility and surface tension to SI values, we first need understand how these quantities scale with the CG degree. Following F\"uchslin~\cite{Fuchslin_JCP_2009}, we note that the $\kBT$ varies linearly with $\Nm$. Since $\rc\sim\Nm^{1/3}$, it follows that $\kappa^{-1,\rm{real}}=\kappa^{-1} \kTc/\rc^3 \sim 1$ is scale-invariant. However, $\sigma^{\rm{real}}=\sigma \kTc/\rc^2 \sim \Nm^{1/3}$. We elaborate further on these issues in a different publication \footnote{Vanya, Sharman, and Elliott, \href{https://arxiv.org/abs/1805.04556}{arxiv:1805.04565}}.

The resulting bulk modulus, which is the inverse of the compressibility, is about three times the experimental value (2.15$\times 10^9$~Pa) and the surface tension is about twice as high as the real value for water (71.5 mN/m). Compared with more precise, atomistically resolved water models such as SPC, which yield about 50~mN/m~\cite{Vega_JCP_2007}, this is not an unreasonable agreement, so we can say that these interaction parameters yield meaningful, if not accurate quantities of interest. However, we now show that there is space for fine-tuning, which would achieve considerably improved precision.

\begin{table}
\begin{ruledtabular}
\begin{tabular}{l|cc}
$\Nm=3$            & Reduced units & Real units \\\hline
Length scale $\rc$ & 1             & 0.818 nm \\
Density            & 6.09          & 997 kg/m$^3$ \\
Surface tension    & 7.01          & 130 mN/m \\
Bulk modulus       & 294           & 6.67$\times 10^9$ Pa \\
\end{tabular}
\end{ruledtabular}
\caption{Predicted physical properties of a typical MDPD liquid with configuration $(A,\,B,\,\rd)=(-40,\,25,\,0.75)$. These can be compared with experimental values $2.15\times10^9$ Pa and $71.5$~mN/m for bulk modulus and surface tension, respectively.}
\label{tbl:real_liq}
\end{table}

\begin{table}
\centering
\begin{ruledtabular}
\begin{tabular}{cccccc}
$\Nm$ & $\rho$ & $A$ & $B$ & $\sigma^{\rm{real}}$ (mN/m) &
$\kappa^{-1,\rm{real}}$ ($10^9$ Pa) \\\hline
1 & 9.99 & $-18.5$ & 3.9 & 71.6 & 2.23 \\
2 & 8.63 & $-18.1$ & 4.9 & 71.5 & 2.16 \\
3 & 7.76 & $-18.2$ & 6.0 & 71.5 & 2.19 \\
4 & 7.23 & $-18.2$ & 6.9 & 71.3 & 2.22 \\
5 & 6.94 & $-18.0$ & 7.4 & 71.4 & 2.20 \\
6 & 6.70 & $-17.9$ & 7.9 & 71.6 & 2.20 \\
7 & 6.55 & $-17.7$ & 8.2 & 71.5 & 2.18 \\
8 & 6.39 & $-17.6$ & 8.6 & 71.4 & 2.18 \\
9 & 6.23 & $-17.6$ & 9.1 & 71.5 & 2.20 \\
10 & 6.12 & $-17.5$ & 9.4 & 71.5 & 2.20 \\
\end{tabular}
\end{ruledtabular}
\caption{Interaction parameters for water at $\rd=0.75$ for a range of CG degrees derived from the fits of density, surface tension and compressibility.}
\label{tbl:water}
\end{table}

\begin{table*}
\centering
\begin{minipage}[b]{0.48\linewidth}
\begin{ruledtabular}
\begin{tabular}{cccccc}
\multicolumn{2}{c}{Ethanol} &&&& \\
\cline{1-2}
$\Nm$ & $\rho$ & $A$ & $B$ & $\sigma^{\rm{real}}$ (mN/m) &
$\kappa^{-1,\rm{real}}$ ($10^9$ Pa) \\\hline
1 & 6.63 & $-20.9$ & 9.7 & 22.3 & 0.84 \\
2 & 5.86 & $-20.3$ & 12.4 & 22.3 & 0.84 \\
3 & 5.49 & $-19.9$ & 14.2 & 22.3 & 0.85 \\
4 & 5.31 & $-19.5$ & 15.2 & 22.3 & 0.84 \\
5 & 5.16 & $-19.2$ & 16.1 & 22.3 & 0.84 \\
\end{tabular}
\end{ruledtabular}
\end{minipage}
\quad
\begin{minipage}[b]{0.48\linewidth}
\begin{ruledtabular}
\begin{tabular}{cccccc}
\multicolumn{2}{c}{Benzene} &&&& \\
\cline{1-2}
$\Nm$ & $\rho$ & $A$ & $B$ & $\sigma^{\rm{real}}$ (mN/m) &
$\kappa^{-1,\rm{real}}$ ($10^9$ Pa) \\\hline
1 & 6.17 & $-33.3$ & 19.6 & 28.0 & 1.05 \\
2 & 5.48 & $-32.3$ & 25.2 & 28.0 & 1.05 \\
3 & 5.18 & $-31.4$ & 28.3 & 28.0 & 1.05 \\
4 & 5.00 & $-30.8$ & 30.7 & 28.0 & 1.05 \\
5 & 4.87 & $-30.3$ & 32.6 & 28.0 & 1.05 \\
\end{tabular}
\end{ruledtabular}
\end{minipage}
\caption{Interaction parameters for ethanol and benzene at $\rd=0.75$ for a range of CG degrees derived from the fits of density, surface tension and compressibility.}
\label{tbl:etoh}
\end{table*}

Usually, in simulating new materials, one desires to go the opposite way, that is start from experimental data and obtain the interaction parameters in reduced units to prepare the material for simulation. Having four equations of four unknowns for the compressibility (eq.~\eqref{eq:kappa}), cutoff $\rc$ (eq.~\eqref{eq:rc}), density $\rho(A,B)$ (eq.~\eqref{eq:rho_fit}) and surface tension $\sigma(A,B)$ (Table.~\ref{tbl:sigma_coeffs}), we can solve these numerically to obtain $A$ and $B$. With resolution $\Delta A=0.1$, $\Delta B=0.1$, it is possible to search through the whole parameter space in reasonable time and choose the configuration with the lowest absolute error defined as follows:
\begin{equation}
\rm{Err} = w\left|1-\frac{\sigma}{\sigma_{\rm L}}\right| + 
\left|1-\frac{\kappa^{-1}}{\kappa^{-1}_{\rm L}}\right|,
\label{eq:err}
\end{equation}
where $\kappa_{\rm L}$ and $\sigma_{\rm L}$ are experimental compressibility and surface tension, respectively, and parameter $w=5$ was chosen to put more weight on the contribution due to the surface tension.

The resulting parameters $A,B$ for water for $\rd=0.75$ are summarised in~\ref{tbl:water}. At CG degrees $\Nm=1$ and 2 the density is relatively high, which implies poor simulation efficiency, but other options yield more reasonable values as well as accurate liquid representations. To demonstrate the robustness of this parametrisation method, Table~\ref{tbl:etoh} shows derived interaction parameters for ethanol and benzene, respectively, as examples of chemically different solvents. These two liquids have several times lower surface tension (22.3~mN/m for ethanol and 28~mN/m for benzene) and compressibility than water, which leads to lower and thus more efficient simulation densities. The Python script to generate these parameters for any chosen liquid and CG degree $\Nm$ and one of the investigated many-body cutoffs $\rd$ is provided in the supplementary material~\footnote{See Supplementary Material at [URL will be inserted by publisher] for a Python script to generate MDPD parameters for any chosen liquid and CG degree $\Nm$ and one of the investigated many-body cutoffs $\rd$.}.

\section{Conclusion}
\label{sec:concl}
In this work we have demonstrated the richness of many-body dissipative particle dynamics and established its suitability for simulating a wide range of mesoscale systems. By systematic variation of the force field parameters we uncovered the regions of liquid, gas and solid phase. We identified one thermodynamically stable solid phase with the bcc lattice, but lacking the proper stress-strain relation. For the liquid phase, we fitted the density and surface tension as a function of the force field parameters and demonstrated how these functional relations can serve to generate the interaction parameters for real liquids. We proved that the resulting top-down parametrisation approach yields reasonable prediction of the force field parameters for water, ethanol and benzene, and in principle can be applied to any other liquid.

This parametrisation enables to apply many-body dissipative particle dynamics to solid/liquid or liquid/gas interfaces of soft matter systems, or porous structures in general. Hence, previously inaccessible environments, such as the catalyst layer of fuel cells, can now be explored~\cite{Weber_JMCA_2014}.

\section{Acknowledgments}
The authors thank Jan Hermann for proofreading the manuscript and Patrick Kiley for helpful discussions about function fitting.
PV, JAE, and PC acknowledge the support of EPSRC. PV and JAE acknowledge the support of Johnson Matthey.

\newpage
\appendix*
\section{Fitting}
Here, we describe in more detail the fitting procedure for the densities and surface tensions as functions of interaction parameters $A,B$ discussed in the main paper. For all the fits, we used the function \texttt{curve\_fit} from the Scipy library~\footnote{\url{https://docs.scipy.org/doc/scipy/
reference/generated/scipy.optimize.curve_fit.html}}.

\subsection{Density profiles}
\begin{figure}
\centering
\includegraphics[width=0.23\textwidth]{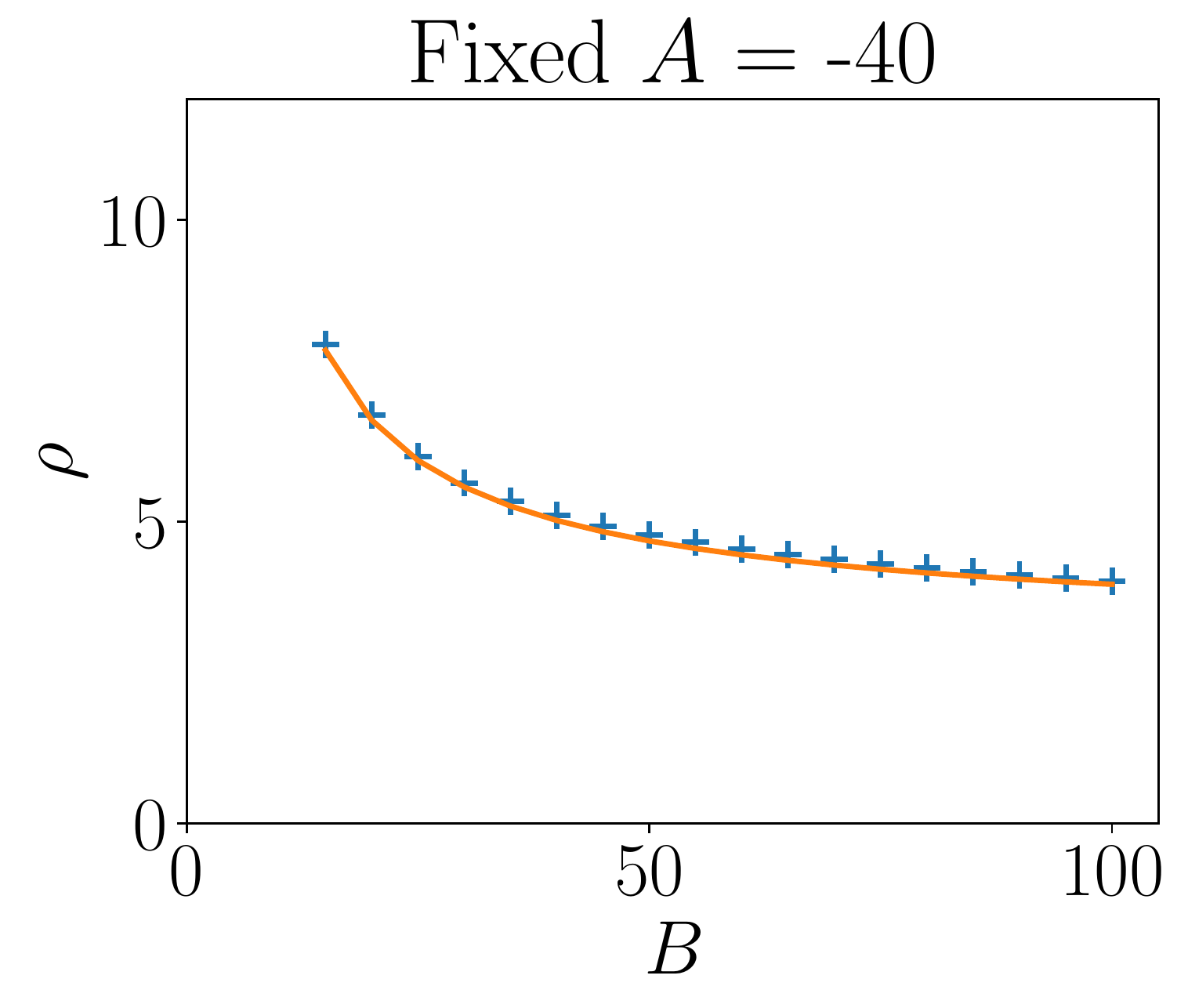}
~
\includegraphics[width=0.23\textwidth]{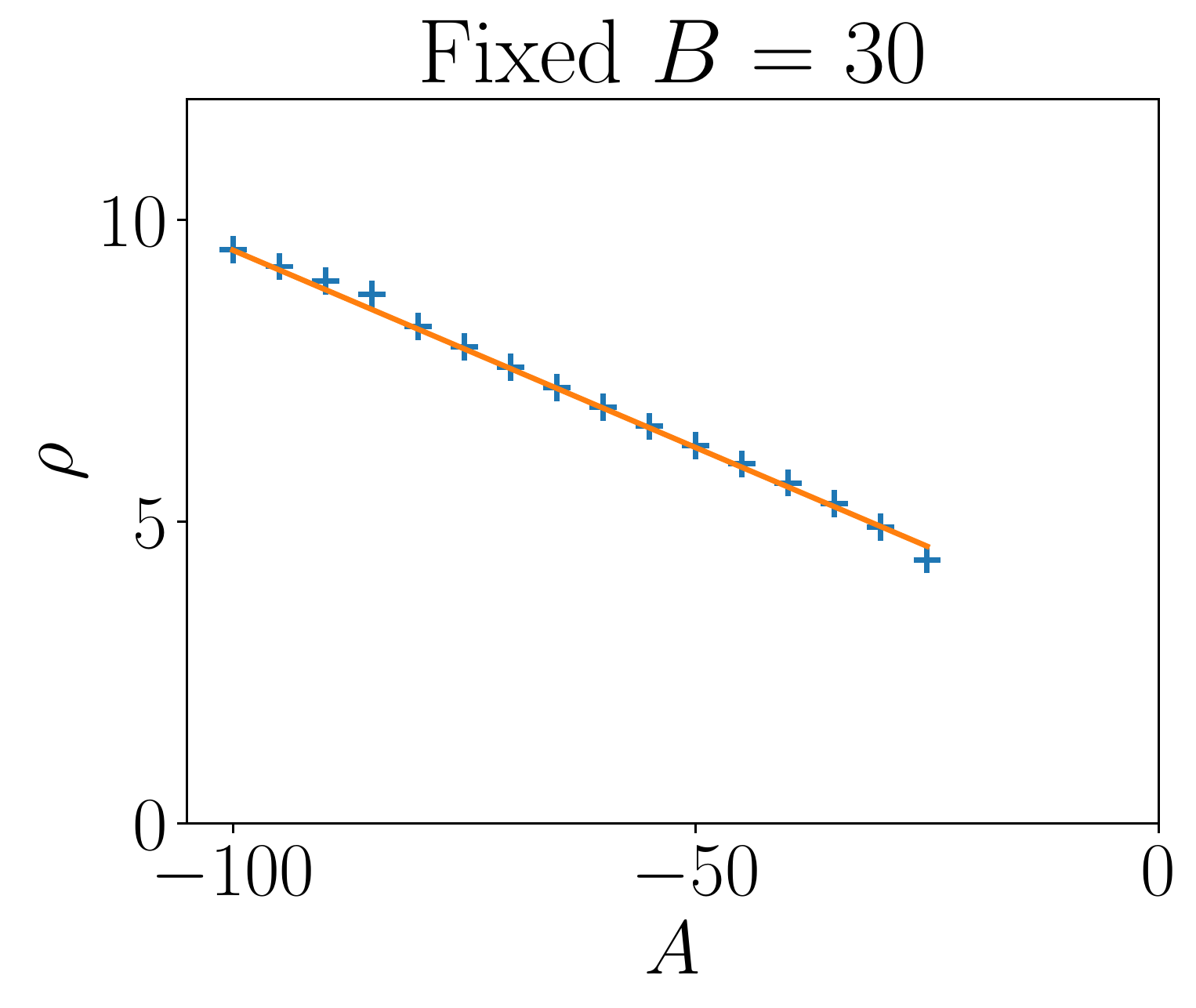}
\caption{Example density surface cuts at $\rd=0.75$ (in reduced units), suggesting linear and power law variation with $A$ and $B$, respectively.}
\label{fig:rho_cuts}
\end{figure}

By visually inspecting the cuts of the density surface $\rho(A,B)$ it is possible to guess several trial functions. At constant $B$, the density varies linearly in the liquid and solid regime for $A<-20$, whereas at constant $A$, the variation follows the power law: $(B-\beta_1)^{\beta_2}$. Example cuts are shown in Fig.~\ref{fig:rho_cuts}.

We applied two versions of the fitting function, composed as the linear combination of the line and the power law, one containing three and the other four parameters. To gauge their relative performance, we randomly split the data into training and validation sets with 80/20 ratio, respectively. We carried out 500 such splits and estimated the average root-mean-square error (RMSE) in the validation set. For further certainty, we also computed the median RMSE to verify that the distribution of the RMSEs is normal. This turned out to be the the case, which was marked by the similar values of median and average RMSE.

The results shown in Table~\ref{tbl:rho_func} reliably conclude that the four-parameter fit performs better for all of the many-body cutoffs $\rd$. However, considering the similarity of the RMSEs and the fact that later, in Section IV of the paper, we would use this fit for deriving the interaction parameters $A,B$ via minimisation, we decided to proceed with the three-parameter fit. The parameters for each of the explored values of $\rd$ are summarised in Table~\ref{tbl:rho_coeffs}.

\begin{figure}
\centering
\includegraphics[width=0.23\textwidth]{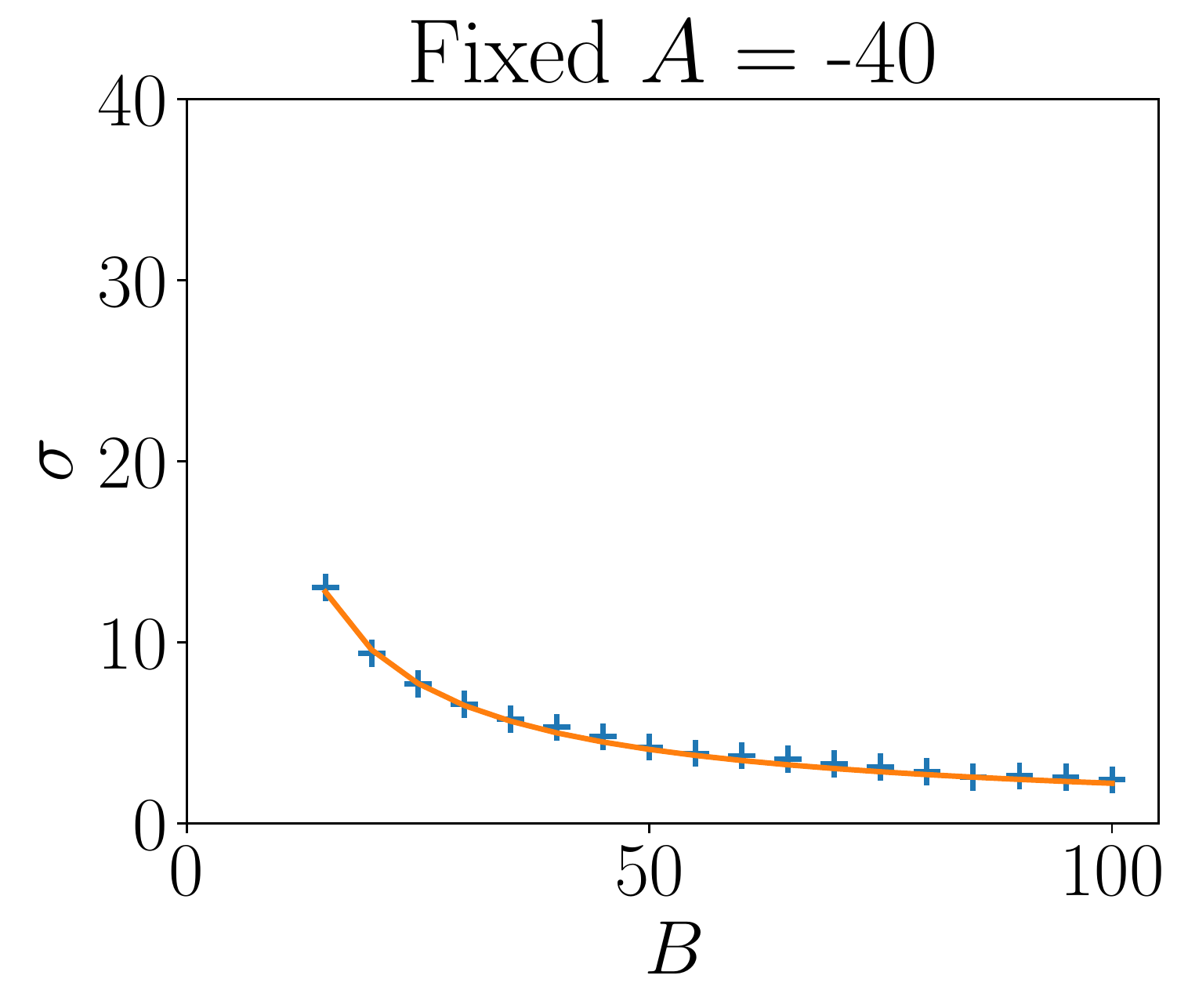}
~
\includegraphics[width=0.23\textwidth]{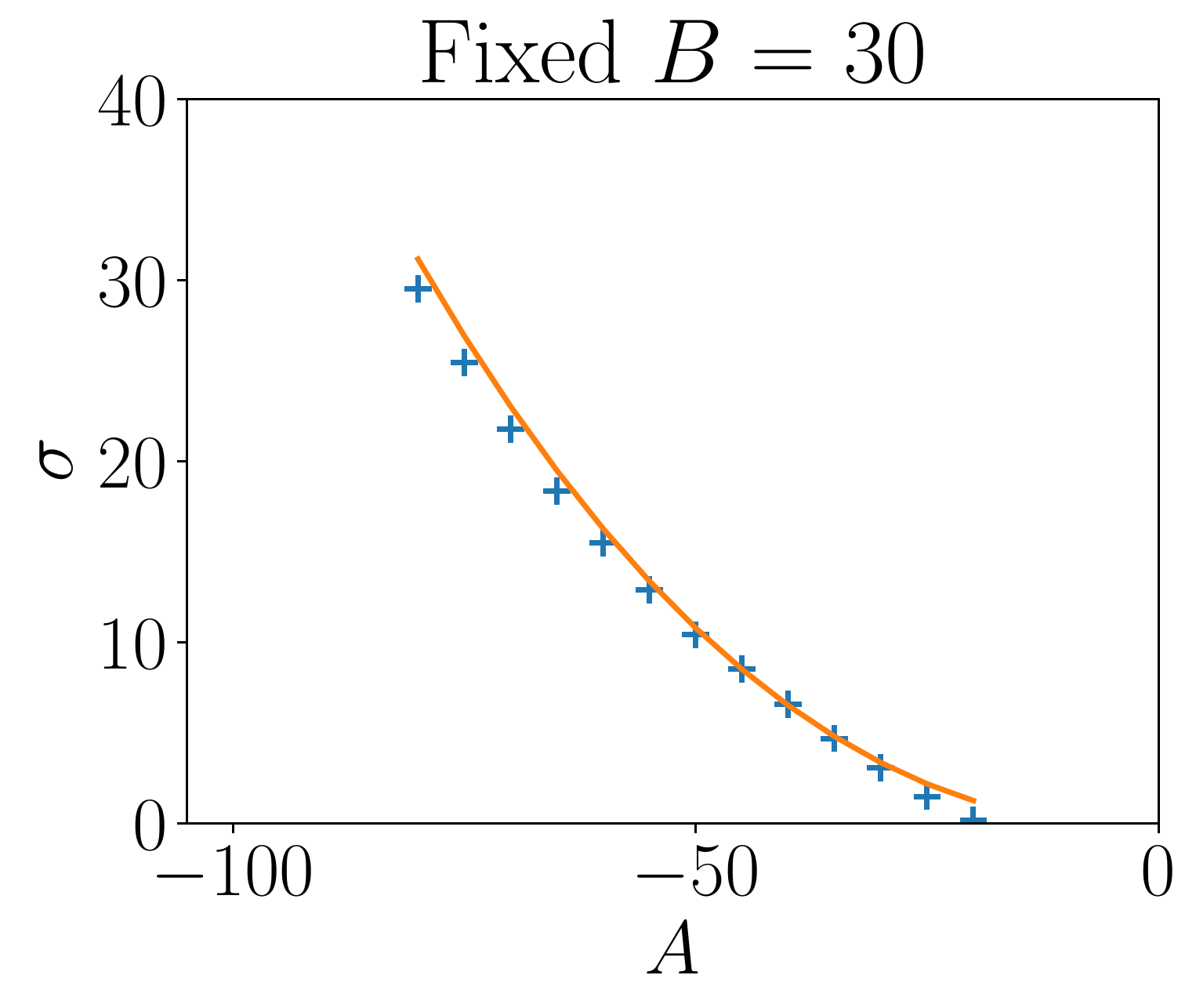}
\caption{Examples of surface tension surface cuts for $\rd=0.75$ (in reduced units).}
\label{fig:sigma_cuts}
\end{figure}

\begin{table}[h!]
\centering
\begin{ruledtabular}
\begin{tabular}{llclll}
& $\rho(A,B)$ & $N_{\rm{param}}$ & \multicolumn{3}{c}{Avg RMSE ($\rd$)}\\\cline{4-6}
&             &                  & 0.65 & 0.75 & 0.85 \\
\hline
1. & $c_1 + c_2(-A)(B-c_3)^{c_4}$ & 4 & 0.20 & 0.16 & 0.13 \\
2. & $c_1 + c_2(-A) B^{c_3}$      & 3 & 0.21 & 0.18 & 0.16 \\
\end{tabular}
\end{ruledtabular}
\caption{Attempted fitting functions for density $\rho(A,B)$ and their respective average RMSEs vs $\rd$'s.}
\label{tbl:rho_func}
\end{table}


\subsection{Surface tension profiles}
Visual inspection of surface tension as a function of $A,B$ (Fig.~\ref{fig:sigma_cuts}) suggests more candidates for fitting functions. The cuts at constant $B$ seemed to indicate a quadratic dependence on $A$, whereas the cuts at constant $A$ gave a power law, as in case of density.

We tried 10 linear combinations of these two functions. In each case, we followed the protocol outlined above: splitting the data 500 times into training and validation sets with 80/20 ratio, and for each split fitting on the training set and computing the RMSE on the points from the validation set. 

The average and median RMSEs showed a non-negligible difference. In such case, we considered median to be a more appropriate measure of the quality of a fitting function. The trial fitting functions and their respective median RMSEs are summarised in Table~\ref{tbl:sigma_func}. Each $\rd$ is best represented by a different function. Deciding between functions with very similar values of median RMSEs, which happened at $\rd=0.85$, we chose the one with the lower number of parameters. The resulting function choices for each value of $\rd$ are summarised in Table~\ref{tbl:sigma_coeffs}.

\begin{table*}[t!]
\centering
\begin{ruledtabular}
\begin{tabular}{llclll}
& $\sigma(A,B)$ & $N_{\rm{param}}$ & \multicolumn{3}{c}{Median RMSE ($\rd$)}\\\cline{4-6}
&               &                  & 0.65 & 0.75 & 0.85 \\
\hline
1. & $(c_1 A^2 + c_2 A + c_3)(B-c_4)^{c_5}$           & 5 & 3.64 & 1.80 & 0.34\\
2. & $(c_1 A^2 + c_2 A + c_3)(B-c_4 + c_5 A)^{c_6}$   & 6 & {\bf 2.09} & NA   & 0.34\\
3. & $(c_1 A^2 + c_2 A + c_3)(B-c_4)^{c_5 + c_6 A}$   & 6 & 3.76 & 1.66 & 0.33\\
4. & $(c_1 A^2 + c_2 A)(B-c_3)^{c_4}$                 & 4 & 3.65 & 1.76 & {\bf 0.34}\\
5. & $(c_1 A^2 + c_2)(B-c_3)^{c_4}$                   & 4 & 3.65 & 1.67 & 0.39\\
6. & $(c_1 A^3 + c_2 A^2 + c_3 A + c_4)(B-c_5)^{c_6}$ & 6 & 3.39 & 1.90 & 0.34\\
7. & $(c_1 A^3 + c_2 A^2 + c_3 A)(B-c_4)^{c_5}$       & 5 & 3.49 & 1.86 & 0.34\\
8. & $(c_1 A^2 + c_2 A)B^{c_3}$                       & 3 & 3.91 & 1.74 & 0.51\\
9. & $(c_1 A^2 + c_2 A + c_3)B^{c_4}$                 & 4 & 3.88 & 1.78 & 0.51\\
10. & $(c_1 A^2 + c_2 A)(B + c_3 A)^{c_4}$            & 4 & 2.33 & {\bf 1.47} & 0.43\\
\end{tabular}
\end{ruledtabular}
\caption{Attempted fitting functions for surface tension $\sigma(A,B)$ and their respective median RMSEs vs $\rd$. The numbers in bold point at the best-fitting functions given the number of parameters.}
\label{tbl:sigma_func}
\end{table*}


\bibliography{ref.bib}
\end{document}